\documentclass[useAMS,usenatbib]{mn2e}
\usepackage[letterpaper,margin=0.65in]{geometry}
\usepackage[ansinew]{inputenc}
\usepackage{amsfonts}
\usepackage{amsmath}
\usepackage{graphicx}
\usepackage{epsfig}
\usepackage{color}
\usepackage{amssymb}
\usepackage[normalem]{ulem}
\usepackage{natbib}
\usepackage{url}
\usepackage{placeins}
\usepackage{times}

\newcommand{\eg}{{e.g., }}
\newcommand{\ie}{{i.e., }}

\title[A Hybrid Approach to Star-Galaxy Classification]{A Hybrid Ensemble Learning Approach to Star-Galaxy Classification}

\author[E. J. Kim, R. J. Brunner \& M. Carrasco Kind]{
  Edward J.~Kim$^1$\thanks{jkim575@illinois.edu},
  Robert J.~Brunner$^{2,3,4}$,
  and Matias Carrasco Kind$^{2,4}$\\
$^1$Department of Physics, University of Illinois, Urbana, IL 61801 USA\\
$^2$Department of Astronomy, University of Illinois, Urbana, IL 61801 USA\\
$^3$Department of Statistics, University of Illinois, Champaign, IL 61820 USA\\
$^4$National Center for Supercomputing Applications, Urbana, IL 61801 USA}

\begin{document}

\date{\today}

\pagerange{\pageref{firstpage}--\pageref{lastpage}} \pubyear{2015}

\maketitle

\label{firstpage}
\begin{abstract}
There exist a variety of star-galaxy classification techniques,
each with their own strengths and weaknesses.
In this paper, we present a novel meta-classification
framework that combines and fully exploits different techniques
to produce a more robust star-galaxy classification.
To demonstrate this hybrid, ensemble approach,
we combine a purely morphological classifier,
a supervised machine learning method based on random forest,
an unsupervised machine learning method based on self-organizing maps,
and a hierarchical Bayesian template fitting method.
Using data from the CFHTLenS survey,
we consider different scenarios:
when a high-quality training set is available with spectroscopic labels from
DEEP2, SDSS, VIPERS, and VVDS, and
when the demographics of sources in a low-quality training set
do not match the demographics of objects in the test data set.
We demonstrate that our Bayesian combination technique improves
the overall performance over any individual classification method
in these scenarios.
Thus, strategies that combine the predictions of different classifiers
may prove to be optimal in currently ongoing and forthcoming
photometric surveys,
such as the Dark Energy Survey and the Large Synoptic Survey Telescope.

\end{abstract}


\section{Introduction}
  \label{section:introduction}

The problem of source classification is fundamental to astronomy
and goes as far back as \citet{messier1781catalogue}.
A variety of different strategies have been developed 
to tackle this long-standing problem,
and yet there is no consensus on
the optimal star-galaxy classification strategy.
The most commonly used method to classify stars and galaxies
in large sky surveys is the morphological separation
~\citep{sebok1979optimal, kron1980photometry, valdes1982resolution,
yee1991faint, vasconcellos2011decision,
henrion2011bayesian}.
It relies on the assumption that
stars appear as point sources
while galaxies appear as resolved sources.
However,
currently ongoing and upcoming large photometric surveys,
such as the Dark Energy Survey
(DES\footnote{http://www.darkenergysurvey.org/})
and the Large Synoptic Survey Telescope
(LSST\footnote{http://www.lsst.org/lsst/}),
will detect a vast number of unresolved galaxies
at faint magnitudes.
Near a survey's limit, the photometric observations
cannot reliably separate stars from unresolved galaxies
by morphology alone without leading to
incompleteness and contamination in the star and galaxy samples.

The contamination of unresolved galaxies can be mitigated
by using training based algorithms.
Machine learning methods have the advantage that
it is easier to include extra information,
such as concentration indices, shape information,
or different model magnitudes.
However,
they are only reliable within the limits of the training data,
and it can be difficult to extrapolate these algorithms
outside the parameter range of the training data.
These techniques can be further categorized into
supervised and unsupervised learning approaches.

In supervised learning, 
the input attributes (\eg magnitudes or colors)
are provided along with the truth labels (\eg star or galaxy).
\citet{odewahn1992automated} pioneered
the application of neural networks
to the star-galaxy classification problem,
and it has become a core part of
the astronomical image processing software
\textsc{SExtractor}~\citep{bertin1996sextractor}.
Other successfully implemented examples include
decision trees~\citep{weir1995automated, suchkov2005census, ball2006robust,
sevilla2015effect}
and Support Vector Machines~\citep*{Fadely2012}.
Unsupervised machine learning techniques
are less common,
as they do not utilize the truth labels during the training process,
and only the input attributes are used.

Physically based template fitting methods
have also been used for the star-galaxy classification
problem~\citep{robin2007stellar, Fadely2012}.
Template fitting approaches infer a source's properties
by finding the best match between
the measured set of magnitudes (or colors)
and the synthetic set of magnitudes (or colors)
computed from a set of spectral templates.
Although it is not necessary to obtain
a high-quality spectroscopic training sample,
these techniques do require
a representative sample of theoretical or empirical templates
that span the possible spectral energy distributions (SEDs)
of stars and galaxies.
Furthermore, they are not exempt from uncertainties
due to measurement errors on the filter response curves,
or from mismatches between the observed magnitudes
and the template SEDs.

In this paper,
we present a novel star-galaxy classification framework
that combines and fully exploits different classification techniques
to produce a more robust classification.
In particular,
we show that the combination of a morphological separation method,
a template fitting technique, a supervised machine learning method,
and an unsupervised machine learning algorithm
can improve the overall performance over any individual method.
In Section~\ref{section:classification_methods},
we describe each of the star-galaxy classification methods.
In Section~\ref{section:classification_combination_methods},
we describe different classification combination techniques.
In Section~\ref{section:data},
we describe the Canada-France Hawaii Telescope Lensing Survey (CFHTLenS)
data set with which we test the algorithms.
In Section~\ref{section:results_and_discussion},
we compare the performance of our combination techniques
to the performance of the individual classification techniques.
Finally, we outline our conclusions in Section~\ref{section:conclusions}.

\section{Classification Methods}
  \label{section:classification_methods}

In this section, we present
four distinct star-galaxy classification techniques.
The first method is a morphological separation method,
which uses a hard cut in the half-light radius vs.\ magnitude plane.
The second method is
a supervised machine learning technique named TPC
(Trees for Probabilistic Classification),
which uses prediction trees and a random forest~\citep{carrascokind2013tpz}.
The third method is
an unsupervised machine learning technique named SOMc,
which uses self-organizing maps (SOMs) and a random atlas
to provide a classification~\citep{carrascokind2014somz}.
The fourth method is
a Hierarchical Bayesian (HB) template fitting technique
based on the work by \citet{Fadely2012},
which fits SED templates from star and galaxy libraries
to an observed set of measured flux values.

Collectively, these four methods represent the majority of all
standard star-galaxy classification approaches published in the literature.
It is very likely that any new classification technique would be
functionally similar to one of these four methods.
Therefore, any of these four methods could in principle be replaced by a similar method.

\subsection{Morphological Separation}

\begin{figure}
  \centering
  \includegraphics[width=\columnwidth]{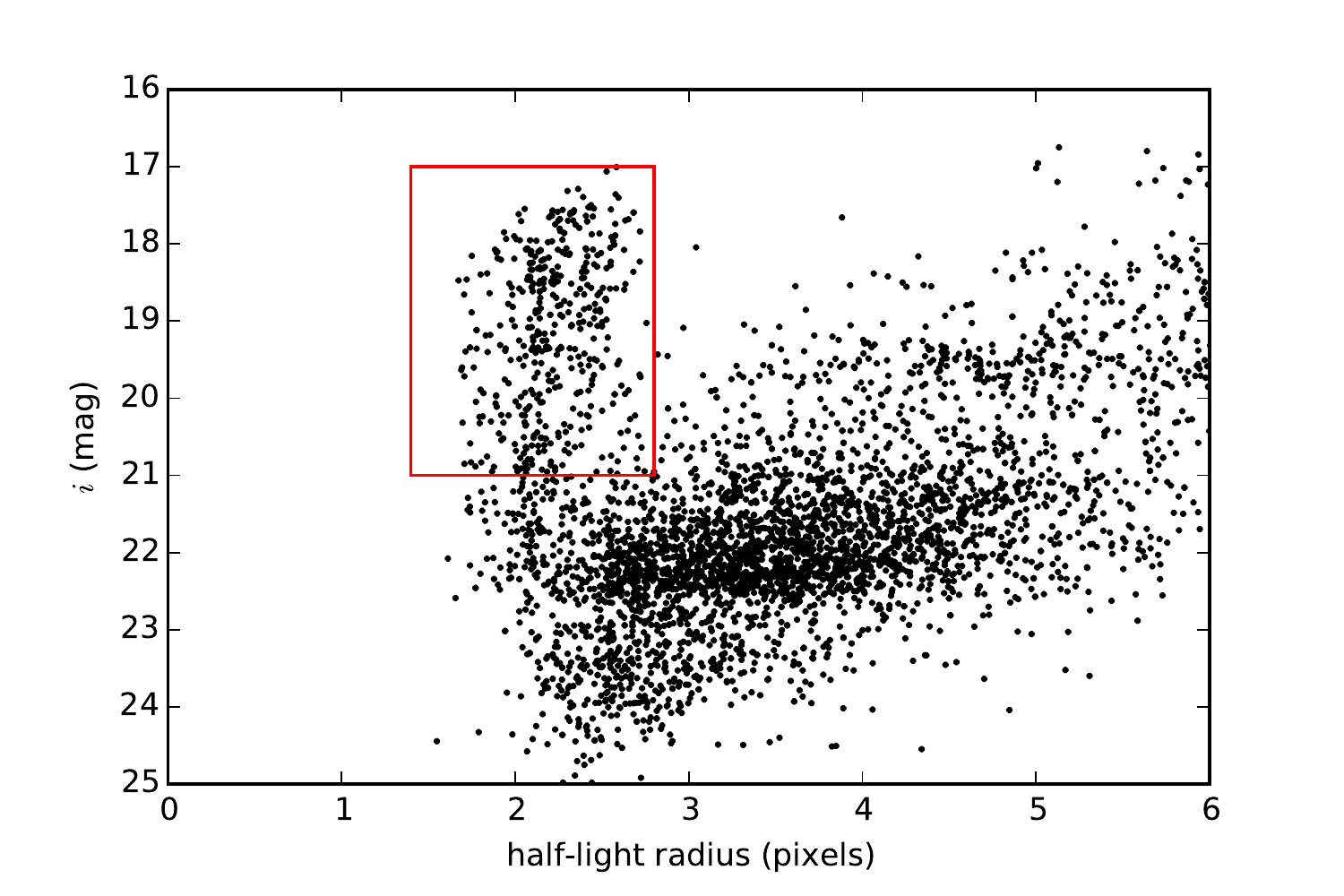}
  \caption{Half-light radius vs.\ magnitude.}
  \label{fig:morph}
\end{figure}

The simplest and perhaps the most widely used approach
to star-galaxy classification is
to make a hard cut in the space of photometric attributes.
As a first-order morphological selection of point sources,
we adopt a technique that is popular among the weak lensing 
community~\citep*{Kaiser1995}.
As Figure~\ref{fig:morph} shows, there is a distinct locus
produced by point sources in the half-light radius
(estimated by \textsc{SExtractor}'s FLUX\_RADIUS parameter)
vs.\ the $i$-band magnitude plane.
A rectangular cut in this size-magnitude plane separates point sources,
which are presumed to be stars,
from resolved sources,
which are presumed to be galaxies.
The boundaries of the selection box are determined by
manually inspecting the size-magnitude diagram.

One of the disadvantages of such cut-based methods is
that it classifies every source with absolute certainty.
It is difficult to justify such a decisive classification
near a survey's magnitude limits,
where measurement uncertainties generally increase.
A more informative approach is to provide probabilistic classifications.
Although a recent work by \citet{henrion2011bayesian}
implemented a probabilistic classification using a Bayesian approach
on the morphological measurements alone,
here we use a cut-based morphological separation
to demonstrate the advantages of our combination techniques.
In particular, we later show that
the binary outputs (\ie 0 or 1) of cut-based methods
can be transformed into
probability estimates by combining them
with the probability outputs from other
probabilistic classification techniques,
such as TPC, SOMc, and HB.

\subsection{Supervised Machine Learning: TPC}

TPC is a parallel, supervised machine learning algorithm
that uses prediction trees and random forest 
techniques~\citep{breiman1984classification, breiman2001random}
to produce a star-galaxy classification.
TPC is a part of a publicly available software package called
\textsc{MLZ}\footnote{http://lcdm.astro.illinois.edu/code/mlz.html}
(Machine Learning for Photo-$z$).
The full software package includes:
\textsc{TPZ}, a supervised photometric redshift (photo-$z$)
estimation technique
\citep[regression mode;][]{carrascokind2013tpz};
TPC, a supervised star-galaxy classification technique
(classification mode);
\textsc{SOM}$z$, an unsupervised photo-$z$ technique
\citep[regression mode;][]{carrascokind2014somz};
and
SOMc, an unsupervised star-galaxy classification technique
(classification mode).

TPC uses classification trees,
a type of prediction trees that are designed to
provide a classification or predict a discrete category.
Prediction trees are built by asking a sequence of questions
that recursively split the data into branches
until a terminal leaf is created
that meets a stopping criterion
(\eg a minimum leaf size).
The optimal split dimension is decided by
choosing the attribute that maximizes
the \textit{Information Gain} ($I_G$), which is defined as

\begin{equation} \label{eq:information_gain}
  I_G \left(D_{\rmn{node}}, X\right)
  = I_d \left( D_{\rmn{node}} \right)
  - \sum_{x \in \rmn{values}(X)}
  \frac{|D_{\rmn{node}, x}|}{|D_{\rmn{node}}|}
  I_d \left( D_{\rmn{node}, x} \right),
\end{equation}

\noindent
where $D_{\rmn{node}}$ is the training data in a given node,
$X$ is one of the possible dimensions (\eg magnitudes or colors)
along which the node is split, and
$x$ are the possible values of a specific dimension $X$.
$|D_{\rmn{node}}|$ and $|D_{\rmn{node}, x}|$ are the size of the total training data
and the number of objects in a given subset $x$ within the current node,
respectively.
$I_d$ is the impurity degree index,
and TPC can calculate $I_d$
from any of the three standard different impurity indices:
\textit{information entropy}, \textit{Gini impurity},
and \textit{classification error}.
In this work, we use the information entropy,
which is defined similarly to the thermodynamic entropy:

\begin{equation}
  I_d \left( D \right)
  = - f_g \log_{2} f_g - \left(1 - f_g\right) \log_{2} \left(1 - f_g\right),
\end{equation}

\noindent
where $f_g$ is the fraction of galaxies in the training data.
At each node in our tree,
we scan all dimensions to identify the split point that
maximizes the information gain as defined by Equation~\ref{eq:information_gain},
and select the attribute that maximizes the impurity index overall.

In a technique called random forest,
we create bootstrap samples
(\ie $N$ randomly selected objects with replacement)
from the input training data
by sampling repeatedly from the magnitudes and colors
using their measurement errors.
We use these bootstrap samples to construct
multiple, uncorrelated prediction trees
whose individual predictions are aggregated to produce
a star-galaxy classification for each source.

We also use a cross-validation technique called
Out-of-Bag~\citep[OOB;][]{breiman1984classification, carrascokind2013tpz}.
When a tree (or a map) is built in TPC (or SOMc),
a fraction of the training data, usually one-third,
is left out and not used in training the trees (or maps).
After a tree is constructed using two-thirds of the training data,
the final tree is applied to the remaining one-third
to make a classification.
This process is repeated for every tree,
and the predictions from each tree are aggregated
for each object to make the final star-galaxy classification.
We emphasize that if an object is used for training a given tree,
it is never used for subsequent prediction by that tree.
Thus, the OOB data is an unbiased estimation of the errors
and can be used as cross-validation data
as long as the OOB data remain similar to the final test data set.
The OOB technique can also provide extra information such as
a ranking of the relative importance of the input attributes
used in the prediction.
The OOB technique can prove extremely valuable
when calibrating the algorithm,
when deciding which attributes to incorporate in the construction of the trees,
and when combining this approach with other techniques.

\subsection{Unsupervised Machine Learning: SOMc}

A self-organizing map~\citep{kohonen1990self, kohonen2001self}
is an unsupervised, artificial neural network algorithm
that is capable of projecting high-dimensional input data
onto a low-dimensional map
through a process of competitive learning.
In astronomical applications,
the high-dimensional input data can be
magnitudes, colors, or some other photometric attributes.
The output map is usually chosen to be two-dimensional
so that the resulting map can be used for visualizing
various properties of the input data.
The differences between a SOM and
typical neural network algorithms are
that a SOM is unsupervised,
there are no hidden layers and therefore no extra parameters,
and it produces a direct mapping
between the training set and the output network.
In fact, a SOM can be viewed as a non-linear generalization
of a principal component analysis (PCA) algorithm~\citep{yin2008self}.

The key characteristic of SOM is that
it retains the topology of the input training set,
revealing correlations between input data that are not obvious.
The method is unsupervised:
the user is not required to specify the desired output
during the creation of the lower-dimensional map,
and the mapping of the components from the input vectors
is a natural outcome of the competitive learning process.

During the construction of a SOM,
each node on the two-dimensional map is represented by
weight vectors of the same dimension
as the number of attributes used to create the map itself.
In an iterative process,
each object in the input sample is individually used
to correct these weight vectors.
This correction is determined so that the specific neuron (or node),
which at a given moment best represents the input source,
is modified along with the weight vectors
of that node's neighboring neurons.
As a result, this sector within the map
becomes a better representation of the current input object.
This process is repeated for every object in the training data,
and the entire process is repeated for several iterations.
Eventually, the SOM converges to its final form where
the training data is separated into groups of similar features.
Although the spectroscopic labels are not used at all
in the learning process, they are used
(only after the map has been constructed)
to generate predictions for each cell in the resulting two-dimensional map.

In a similar approach to random forest in TPZ and TPC,
SOM$z$ uses a technique called \textit{random atlas}
to provide photo-$z$ estimation \citep{carrascokind2014somz}.
In random atlas, the prediction trees of random forest are replaced by maps,
and each map is constructed from different bootstrap samples
of the training data.
Furthermore, we create random realizations of the training data
by perturbing the magnitudes and colors by their measurement errors.
For each map, we can either use all available attributes,
or randomly select a subsample of the attribute space.
This SOM implementation can also be
applied to the classification problem,
and we refer to it as SOMc
in order to differentiate it from 
the photo-$z$ estimation problem (regression mode).
We also use the random atlas approach
in some of the classification combination approaches as discussed in
Section~\ref{section:classification_combination_methods}.

One of the most important parameter in SOMc
is the topology of the two-dimensional SOM,
which can be rectangular, hexagonal, or spherical.
In our SOM implementation, it is also possible
to use periodic boundary conditions for the non-spherical cases.
The spherical topology is by definition periodic and
is constructed by using
\textsc{HEALPIX}~\citep{gorski2005healpix}.
Similar to TPC,
we use the OOB technique to make an unbiased estimation of errors.
We determine the optimal parameters by performing
a grid search in the parameter space of different toplogies,
as well as other SOM parameters, for the OOB data.
We find that the spherical topology gives the best performance
for the CFHTLenS data, likely due to its natural periodicity.
Thus, we use a spherical topology to classify stars and galaxies
in the CFHTLenS data.
For a complete description of the SOM implementation
and its application to the estimation of
photo-$z$ probability density functions (photo-$z$ PDFs),
we refer the reader to \cite{carrascokind2014somz}.

\subsection{Template fitting: Hierarchical Bayesian}

One of the most common methods to classify a source
based on its observed magnitudes is template fitting.
Template fitting algorithms do not require a spectroscopic training sample;
there is no need for additional knowledge outside the
observed data and the template SEDs.
However, any incompleteness in our knowledge of the template SEDs
that fully span the possible SEDs of observed sources
may lead to misclassification of sources.

Bayesian algorithms use Bayesian inference to quantify
the relative probability that each template matches
the input photometry
and determine a probability estimate by computing
the posterior that a source is a star or a galaxy.
In this work, we have modified and parallelized
a publicly available Hierarchical Bayesian (HB) template fitting
algorithm by \cite{Fadely2012}.
In this section, we provide a brief description
of the HB template fitting technique;
for the details of the underlying HB approach,
we refer the reader to \cite{Fadely2012}.

We write the posterior probability that a source is a star as

\begin{equation} \label{eq:overall_posterior}
P \left( S | \bmath{x}, \bmath{\theta} \right)
= P \left( \bmath{x} | S, \bmath{\theta} \right)
P \left( S | \bmath{\theta} \right),
\end{equation}

\noindent
where $\bmath{x}$ represents a given set of observed magnitudes,.
We have also introduced the \textit{hyperparameter} $\bmath{\theta}$,
a nuisance parameter that characterizes our uncertainty
in the prior distribution.
To compute the likelihood that a source is a star,
we marginalize over all star and galaxy templates $\bmath{T}$.
In a template-fitting approach,
we marginalize by summing up
the likelihood that a source has the set of magnitudes $\bmath{x}$
for a given star template
as well as the likelihood for a given galaxy template:

\begin{equation} \label{eq:marginalize_template}
  P \left(\bmath{x} | S, \bmath{\theta} \right)
  = \sum_{t \in \bmath{T}}
  P \left(\bmath{x} | S, t, \bmath{\theta} \right)
  P \left(t | S, \bmath{\theta} \right).
\end{equation}

\noindent
The likelihood of each template
$P \left( \bmath{x} | S, \bmath{\theta} \right)$
is itself marginalized over the uncertainty
in the template-fitting coefficient.
Furthermore, for galaxy templates, we introduce another step that 
marginalizes the likelihood by redshifting a given galaxy template
by a factor of $1 + z$.

Marginalization in Equation~\ref{eq:marginalize_template}
requires that we specify the prior probability
$P \left(t | S, \bmath{\theta} \right)$
that a source has a spectral template $t$ (at a given redshift).
Thus, the probability that a source is a star (or a galaxy)
is either the posterior probability itself
if a prior is used,
or the likelihood itself if an uninformative prior is used.
In a Bayesian analysis, it is preferable to use a prior,
which can be directly computed either from physical assumptions,
or from an empirical function calibrated by
using a spectroscopic training sample.
In an HB approach, the entire sample of sources is used to
infer the prior probabilities for each individual source.

Since the templates are discrete in both SED shape and physical properties,
we parametrize the prior probability of each template
as a discrete set of weights such that

\begin{equation}
\sum_{t \in \bmath{T}}
P \left(t | S, \bmath{\theta} \right) = 1.
\end{equation}

\noindent
Similarly, we also parametrize the overall prior probability,
$\left(S | \bmath{\theta}\right)$, in Equation~\ref{eq:overall_posterior},
as a weight.
These weights correspond to the hyperparameters,
which can be inferred by sampling
the posterior probability distribution in the hyperparameter space.
For the sampling, we use \textsc{emcee}, a Python implementation of the
affine-invariant Markov Chain Monte Carlo (MCMC) ensemble sampler~\citep{Foreman-Mackey2013}.  

As the goal of template fitting methods is to minimize
the difference between observed and theoretical magnitudes,
this approach heavily relies on
both the use of SED templates and
the accuracy of the transmission functions
for the filters used for particular survey.
For our stellar templates,
we use the empirical SED library from \citet{pickles1998stellar}.
The Pickles library consists of 131 stellar templates,
which span all normal spectral types
and luminosity classes at solar abundance,
as well as metal-poor and metal-rich F--K dwarf 
and G--K giant and supergiant stars.
We supplement the stellar library with
100 SEDs from \citet{chabrier2000evolutionary},
which include low mass stars and brown dwarfs
with different $T_{\rmn{eff}}$ and surface gravities.
We also include four white dwarf templates of
\citet*{bohlin1995white}, for a total of 235 templates
in our final stellar library.  
For our galaxy templates,
we use four CWW spectra from \cite*{coleman1980colors},
which include an Elliptical, an Sba, an Sbb,
and an Irregular galaxy template.
When extending an analysis to higher redshifts,
the CWW library is often augmented with
two star bursting galaxy templates from \cite{kinney1996template}.
From the six original CWW and Kinney spectra,
intermediate templates are created by interpolation,
for a total of 51 SEDs in our final galaxy library.

All of the above templates are convolved
with the filter response curves to generate model magnitudes.
These response curves consist of
$u$, $g$, $r$, $i$, $z$ filter transmission functions
for the observations taken by the
Canada-France Hawaii Telescope (CFHT).

\section{Classification combination methods}
  \label{section:classification_combination_methods}

Building on the work in the field of ensemble learning,
we combine the predictions from
individual star-galaxy classification techniques
using four combination techniques.
The main idea behind ensemble learning is to weight
the predictions from individual models
and combine them to obtain a prediction
that outperforms every one of 
them individually~\citep{rokach2010ensemble}.

\subsection{Unsupervised Binning}
  \label{section:random_atlas}

Given the variety of star-galaxy classification methods
we are using,
we fully expect the relative performance
of the individual techniques to vary across
the parameter space spanned by the data.
For example, it is reasonable to expect 
supervised techniques to outperform other techniques
in areas of parameter space that are well-populated
with training data.
Similarly, we can expect unsupervised approaches
such as SOM or template fitting approaches 
to generally perform better when a training sample
is either sparse or unavailable.

We therefore adopt a binning strategy similar to 
\cite{carrascokind2014exhausting}.
In this binning strategy,
we allow different classifier combinations 
in different parts of parameter space
by creating two-dimensional SOM representations of
the full nine-dimensional magnitude-color space:
$u$, $g$, $r$, $i$, $z$, $u-g$, $g-r$, $r-i$, and $i-z$.
A SOM representation can be rectangular, hexagonal, or spherical;
here we choose a 10$\times$10 rectangular topology to facilitate 
visualization as shown in Figure~\ref{fig:som_colors}.
We note that this choice is mainly for convenience
and that the optimal topology and map size would likely depend on
a number of factors, such as the number of objects and attributes.
For all combination methods,
we use only the OOB (cross-validation) data contained in each cell
to compute the relative weights for the base classifiers.
The weights within individual cells are then applied to
the blind test data set to make the prediction.

Furthermore, we construct a collection of SOM representations
and subsequently combine the predictions from each map
into a meta-prediction.
Given a training sample of $N$ sources,
we generate $N_R$ random realizations of training data
by perturbing the attributes
with the measured uncertainty for each attribute.
The uncertainties are assumed to be normally distributed.
In this manner,
we reduce the bias towards the data
and introduce randomness in a systematic manner.
For each random realization of a training sample,
we create $N_M$ bootstrap samples of size $N$
to generate $N_M$ different maps.

After all maps are built,
we have a total of $N_R \times N_M$ probabilistic outputs
for each of the $N$ sources.
To produce a single probability estimate for each source,
we could take the mean, the median, or some other simple statistic.
With a sufficient number of maps,
we find that there is usually negligible difference
between taking the mean and taking the median, and
use the median in the following sections.
We note that it is also possible to establish confidence intervals
using the distribution of the probability estimates.

\subsection{Weighted Average}

The simplest approach to combine different combination techniques is
to simply add the individual classifications from the base classifiers
and renormalize the sum.
In this case, the final probability is given by
\begin{equation}
P\left(S | \bmath{x}, \bmath{M} \right)
= \sum_{i} P \left( S | \bmath{x}, M_{i} \right),
\end{equation}

\noindent
where $\bmath{M}$ is the set of models
(TPC, SOMc, HB, and morphological separation
in our work).
We improve on this simple approach
by using the binning strategy to
calculate the weighted average of objects in each SOM cell
separately for each map,
and then combine the predictions from each map into a final prediction.

\subsection{Bucket of Models (BoM)}

After the multi-dimensional input data have been binned,
we can use the cross-validation data
to choose the best model within each bin,
and use only that model within that specific bin
to make predictions for the test data.
We use the mean squared error
(MSE; also known as Brier score~\citep{brier1950verification})
as a classification error metric. We define MSE as

\begin{equation} \label{eq:mse}
  \rmn{MSE} = \frac{1}{N} \sum^{N - 1}_{i = 0}
  \left( y_i - \hat{y}_i \right)^2,
\end{equation}

\noindent
where $\hat{y}_i$ is the actual truth value
(\eg 0 or 1) of the $i^{\text{th}}$ data, 
and $y_{i}$ is the probability prediction made by the models.
Thus, a model with the minimum MSE is chosen in each bin,
and is assigned a weight of one,
and zero for all other models.
However, the chosen model is allowed to vary between different bins.

\subsection{Stacking}

Instead of selecting a single model that performs best within each bin,
we can train a learning algorithm to combine the output values of
several other base classifiers in each bin.
An ensemble learning method of using a meta-classifier
to combine lower-level classifiers is known as \textit{stacking}
or \textit{stacked generalization}~\citep{wolpert1992stacked}.
Although any arbitrary algorithm
can theoretically be used as a meta-classifier,
a logistic regression or a linear regression is often used
in practice.
In our work,
we use a single-layer multi-response linear regression algorithm,
which often shows the best performance~\citep{breiman1996stacked,
ting1999issues}. 
This algorithm
is a variant of the least-square regression algorithm,
where a linear regression model is constructed for each class.

\subsection{Bayesian Model Combination}

We also use a model combination technique known as
Bayesian Model Combination~\citep[BMC;][]{Monteith2011}, which
uses Bayesian principles to generate an ensemble combination of
different classifiers.
The posterior probability that a source is a star is given by

\begin{equation} \label{eq:p_star_bmc}
  P \left(S | \bmath{x}, \bmath{D}, \bmath{M}, \bmath{E} \right)
  = \sum_{e \in \bmath{E}} P \left(S | \bmath{x}, \bmath{M}, e \right)
  P \left(e | \bmath{D} \right),
\end{equation}

\noindent
where $\bmath{D}$ is the data set,
and $e$ is an element in the ensemble space $\bmath{E}$ of possible model combinations.
By Bayes' Theorem, the posterior probability of $e$ given $\bmath{D}$ is given by

\begin{equation} \label{eq:p_ensemble}
  P \left(e | \bmath{D} \right)
  = \frac{P \left(e \right)}{P \left(\bmath{D} \right)}
  \prod_{d \in \bmath{D}} P \left( d | e \right)
  \propto P \left(e\right) \prod_{d \in \bmath{D}} P \left(d | e \right).
\end{equation}

\noindent
Here, $P\left(e \right)$ is the prior probability of $e$,
which we assume to be uniform.
The product of $P\left(d | e\right)$ is
over all individual data $d$ in the training data $\bmath{D}$,
and $P\left(\bmath{D}\right)$ is merely a normalization factor
and not important.

For binary classifiers whose output is either zero or one
(\eg a cut-based morphological separation),
we assume that each example is corrupted with
an average error rate $\epsilon$.
This means that
$P\left(d|e\right) = 1-\epsilon$ if the combination $e$
correctly predicts class $\hat{y}_i$ for the $i^{\text{th}}$ object,
and $P\left(d|e\right) = \epsilon$ if it predicts an incorrect class.
The average rate $\epsilon$ can be estimated by
the fraction $\left(M_g + M_s\right) / N$,
where $M_g$ is the number of true galaxies classified as stars,
$M_s$ is the number of true stars classified as galaxies,
and $N$ is the total number of sources.
Equation~\ref{eq:p_ensemble} then becomes

\begin{equation} \label{eq:p_ensemble_discrete}
  P \left( e | \bmath{D} \right) \propto 
  P \left( e \right) \left(1 - \epsilon \right)^{N - M_s - M_g}
  \left( \epsilon \right)^{M_s + M_g}.
\end{equation}

\noindent
For probabilistic classifiers,
we can directly use the probabilistic predictions
and write Equation~\ref{eq:p_ensemble} as

\begin{equation} \label{eq:p_ensemble_prob}
  P \left( e | \bmath{D} \right) \propto 
  P \left( e \right) \prod_{i=0}^{N-1}
  \hat{y}_i y_i + 
  \left(1 - \hat{y}_i\right) \left(1 - y_i\right).
\end{equation}

Although the space $\bmath{E}$ of potential model combinations
is in principle infinite,
we can produce a reasonable finite set
of potential model combinations by using sampling techniques.
In our implementation,
the weights of each combination of the base classifiers
is obtained by sampling from a Dirichlet distribution.
We first set all alpha values of a Dirichlet distribution to unity.
We then sample this distribution $q$ times
to obtain $q$ sets of weights.
For each combination,
we assume a uniform prior and
calculate $P\left(e|\bmath{D}\right)$ using
Equation~\ref{eq:p_ensemble_discrete} or \ref{eq:p_ensemble_prob}.
We select the combination with the highest $P\left(e|\bmath{D}\right)$,
and update the alpha values by
adding the weights of the most probable combination
to the current alpha values.
The next $q$ sets of weights are drawn
using the updated alpha values.

We continue the sampling process until
we reach a predefined number of combinations,
and finally use Equation~\ref{eq:p_star_bmc} to compute
the posterior probability that a source is a star (or a galaxy).
In this paper, we use a $q$ value of three,
and 1,000 model combinations are considered.

We also use a binned version of the BMC technique,
where we use a SOM representation
to apply different model combinations
for different regions of the parameter space.
We however note that introducing randomness
though the construction of $N_R \times N_M$ different SOM representations
does not show significant
improvement over using only one single SOM representation.
This similarity is likely due to the randomness 
that has already been introduced by 
sampling from the Dirichlet distribution.
Thus, our BMC technique uses one SOM,
while other base models (WA, BoM, and stacking)
generate $N_R$ random realizations of $N_M$ maps.

\section{Data}
  \label{section:data}

We use photometric data from
the Canada-France-Hawaii Telescope Lensing Survey
\cite[CFHTLenS\footnote{http://www.cfhtlens.org/};][]
{heymans2012cfhtlens,erben2013cfhtlens,hildebrandt2012cfhtlens}.
This catalog consists of more than twenty five million objects
with a limiting magnitude of $i_{\text{AB}} \approx 25.5$. 
It covers a total of 154 square degrees
in the four fields (named W1, W2, W3, and W4)
of CFHT Legacy Survey~\citep[CFHTLS;][]{gwyn2012canada}
observed in the five photometric bands:
$u$, $g$, $r$, $i$, and $z$.

We have cross-matched reliable spectroscopic galaxies from
the Deep Extragalactic Evolutionary Probe Phase 2~
\citep[DEEP2;][]{davis2003science,newman2013deep2},
the Sloan Digital Sky Survey Data Release 10~\citep[SDSS-DR10]{Ahn2014},
the VIsible imaging Multi-Object Spectrograph (VIMOS)
Very Large Telescope (VLT) Deep Survey~
\citep[VVDS;][]{le2005vimos,garilli2008vimos}, and
the VIMOS Public Extragalactic Redshift
Survey~\citep[VIPERS;][]{garilli2014vimos}.
We have selected only sources with very secure
redshifts and no bad flags (quality flags -1, 3, and 4 for DEEP2;
quality flag 0 for SDSS; quality flags 3, 4, 23, and 24 for VIPERS
and VVDS).
In the end, we have 8,545 stars and 57,843 galaxies available
for the training and testing processes.
We randomly select 13,278 objects for the blind testing set,
and use the remainder for training and cross-validation.
While HB uses only the magnitudes in the five bands,
$u$, $g$, $r$, $i$, and $z$,
TPC and SOMc are trained with a total of 9 attributes: the five magnitudes
and their corresponding colors, $u-g$, $g-r$, $r-i$, and $i-z$.
The morphological separation method uses \textsc{SExtractor}'s 
FLUX\_RADIUS parameter provided by the CFHTLenS catalog.

Our goal here is not to obtain the best classifier
performance; for this we would have fine tuned individual base
classifiers and chosen sophisticated models best suited to the
particular properties of the CFHTLenS data.
For example, \cite{hildebrandt2012cfhtlens} suggest that
all objects with $i > 23$ in the CFHTLenS data set
may be classified as galaxies
without significant incompleteness and contamination in the galaxy sample.
Although this approach works because
the high Galactic latitude fields of the CFHTLS contain
relatively few stars,
it is very unlikely that such an approach
will meet the science requirements for the quality of
star-galaxy classification in lower-latitude, star-crowded fields.
Rather, our goal for the CFHTLenS data set
is to demonstrate the usefulness
of combining different classifiers even when
the base classifiers may be poor or trained on partial data.
We also note that the relatively few number of stars
in the CFHTLS fields might paint too positive a picture of completeness and purity,
especially for the stars.
Thus, we caution the reader that the specific completeness and purity values
will likely vary in other surveys that observe large portions of the sky,
and we emphasize once again that our aim is to highlight that there is a
relative improvement in performance when we combine multiple star-galaxy
classification techniques to generate a meta-classification.

\section{Results and Discussion}
  \label{section:results_and_discussion}

In this section, we present the classification performance of
the four different combination techniques,
as well as the individual star-galaxy classification techniques
on the CFHTLenS test data.

\subsection{Classification Metrics}

Probabilistic classification models can be considered as
functions that output a probability estimate of each source
to be in one of the classes (\eg a star or a galaxy).
Although the probability estimate can be used as a weight 
in subsequent analyses to improve or enhance
a particular measurement~\citep{ross2011ameliorating},
it can also be converted into a class label
by using a threshold (a probability cut).
The simplest way to choose the threshold is to set it to a fixed value,
\eg $p_\rmn{cut} = 0.5$.
This is, in fact, what is often done
\citep[\eg][]{henrion2011bayesian, Fadely2012}.
However, choosing $0.5$ as a threshold is not the best choice
for an unbalanced data set, where galaxies outnumber stars.
Furthermore, setting a fixed threshold ignores the operating condition 
(\eg science requirements, stellar distribution, misclassification costs)
where the model will be applied.

\subsubsection{Receiver Operating Characteristic Curve}

When we have no information about the operating condition
when evaluating the performance of classifiers,
there are effective tools such as
the Receiver Operating Characteristic (ROC) curve
\citep*{swets2000better}.
An ROC curve is a graphical plot that illustrates the true positive rate
versus the false positive rate of a binary classifier
as its classification threshold is varied.
The Area Under the Curve (AUC) summarizes the curve information
in a single number,
and can be used as an assessment of the overall performance.

\subsubsection{Completeness and Purity}

In astronomical applications,
the operating condition usually translates to
the completeness and purity requirements of the star or galaxy sample.
We define the galaxy \textit{completeness}
$c_g$ (also known as recall or sensitivity) as
the fraction of the number of true galaxies classified as galaxies
out of the total number of true galaxies,
\begin{equation}
c_g = \frac{N_g}{N_g + M_g},
\end{equation}

\noindent
where $N_g$ is the number of true galaxies classified as galaxies,
and $M_g$ is the number of true galaxies classified as stars.
We define the galaxy \textit{purity} $p_g$ (also known as precision
or positive predictive value)
as the fraction of the number of true galaxies classified as galaxies
out of the total number of objects classified as galaxies, 
\begin{equation}
p_g = \frac{N_g}{N_g + M_s},
\end{equation}

\noindent
where $M_s$ is the number of true stars classified as galaxies.
Star completeness and purity are defined in a similar manner.

One of the advantages of a probabilistic classification is
that the threshold can be adjusted to produce
a more complete but less pure sample,
or a less complete but more pure one.
To compare the performance of probabilistic classification techniques
with that of morphological separation,
which has a fixed completeness ($c_g = 0.9964$, $c_s = 0.7145$)
at a certain purity ($p_g = 0.9597$, $p_s = 0.9666$),
we adjust the threshold of probabilistic classifiers
until the galaxy completeness $c_g$ matches
that of morphological separation
to compute the galaxy purity $p_{g}$ at $c_g=0.9964$.
Similarly, the star purity $p_{s}$ at $c_{s}=0.7145$
is computed by adjusting the threshold
until the star completeness of each classifier is equal to
that of morphological separation.

We can also compare the performance of different classification techniques
by assuming an arbitrary operating condition.
For example, weak lensing science measurements
of the DES require $c_g > 0.960$ and $p_g > 0.778$
to control both the statistical and systematic errors
on the cosmological parameters,
and $c_s > 0.250$ and $p_s > 0.970$
for stellar Point Spread Function (PSF) calibration
\citep{soumagnac2013star}.
Although these values will likely be different
for the science cases of the CFHTLenS data,
we adopt these values to compare the classification performance
at a reasonable operating condition.
Thus, we compute $p_{g}$ at $c_g=0.960$
and $p_{s}$ at $c_s=0.250$.
We also use the MSE defined in Equation~\ref{eq:mse}
as a classification error metric.

\begin{table}
  \caption{The definition of the classification performance metrics.}
  \centering
  \begin{tabular}{c l}
  Metric & Meaning \\
  \hline
  AUC & Area under the Receiver Operating Curve \\
  MSE & Mean squared error \\
  $c_g$ & Galaxy completeness \\
  $p_g$ & Galaxy purity \\
  $c_s$ & Star completeness \\
  $p_s$ & Star purity \\
  $p_g(c_g=x)$ & Galaxy purity at $x$ galaxy completeness \\
  $p_s(c_s=x)$ & Star purity at $x$ star completeness \\
  \end{tabular}
  \label{table:metrics}
\end{table}

\subsection{Classifier Combination}
  \label{section:rich_training}

\begin{table*}
  \caption{A summary of the classification performance metrics
           for the four individual methods
           and the four different classification combination methods
           as applied to the CFHTLenS data,
           with no cut applied to the training data set.
           The definition of the metrics is summarized in
           Table~\ref{table:metrics}.
           The bold entries highlight the best performance values
           within each column.
           Note that some objects in the test set have bad or missing
           values (\eg $-99$ or $99$) in one or more attributes,
           which are included here (but are omitted, for example,
           in Figure~\ref{fig:purity_mag_integrated} when the corrsponding
           attribute is not available.)}
  \centering
  \begin{tabular}{l c c c c c c}
  Classifier & AUC & MSE &
  $p_{g}\left(c_g=0.9964\right)$ & $p_{s}\left(c_s=0.7145\right)$ &
  $p_{g}\left(c_g=0.9600\right)$ & $p_{s}\left(c_s=0.2500\right)$ \\
  \hline
  TPC        & \textbf{0.9870} & 0.0208 & 0.9714 & 0.9838 & 0.9918 & 0.9977 \\
  SOMc       & 0.9683 & 0.0452 & 0.9125 & 0.8454 & 0.9788 & 0.9551 \\
  HB         & 0.9403 & 0.0705 & 0.9219 & 0.7017 & 0.9471 & 0.6963 \\
  Morphology & - & 0.0397 & 0.9597 & 0.9666 & - & - \\
  WA         & 0.9806 & 0.0266 & 0.9755 & 0.9926 & 0.9872 & 0.9977 \\
  BoM        & 0.9870 & 0.0208 & 0.9714 & 0.9838 & 0.9918 & 0.9977 \\
  Stacking   & 0.9842 & 0.0194 & 0.9752 & 0.9902 & 0.9918 & \textbf{1.0000} \\
  BMC        & 0.9852 & \textbf{0.0174} & \textbf{0.9800} & $\textbf{0.9959}$ &
  $\textbf{0.9924}$ & \textbf{1.0000} \\
\end{tabular}

  \label{table:metrics_all}
\end{table*}

We present in Table~\ref{table:metrics_all}
the classification performance obtained by
applying the four different combination techniques,
as well as the individual star-galaxy classification techniques,
on the CFHTLenS test data.
The bold entries highlight the best technique for any particular metric.
The first four rows show the performance of four individual star-galaxy
classification techniques.
Given a high-quality training data, it is not surprising that 
our supervised machine learning technique TPC
outperforms other unsupervised techniques.
TPC is thus shown in the first row as the benchmark.

The simplest of the combination techniques, WA and BoM,
generally do not perform better than TPC.
It is also interesting that,
even with binning the parameter space and selecting the best model
within each bin,
BoM almost always chooses TPC as the best model in all bins,
and therefore gives the same performance as TPC in the end.
However, our BMC and stacking techniques have a similar performance
and often outperform TPC.
Although TPC shows the best performance as measured by the AUC,
BMC shows the best performance in all other metrics.

\begin{figure}
  \centering
  \includegraphics[width=\columnwidth]{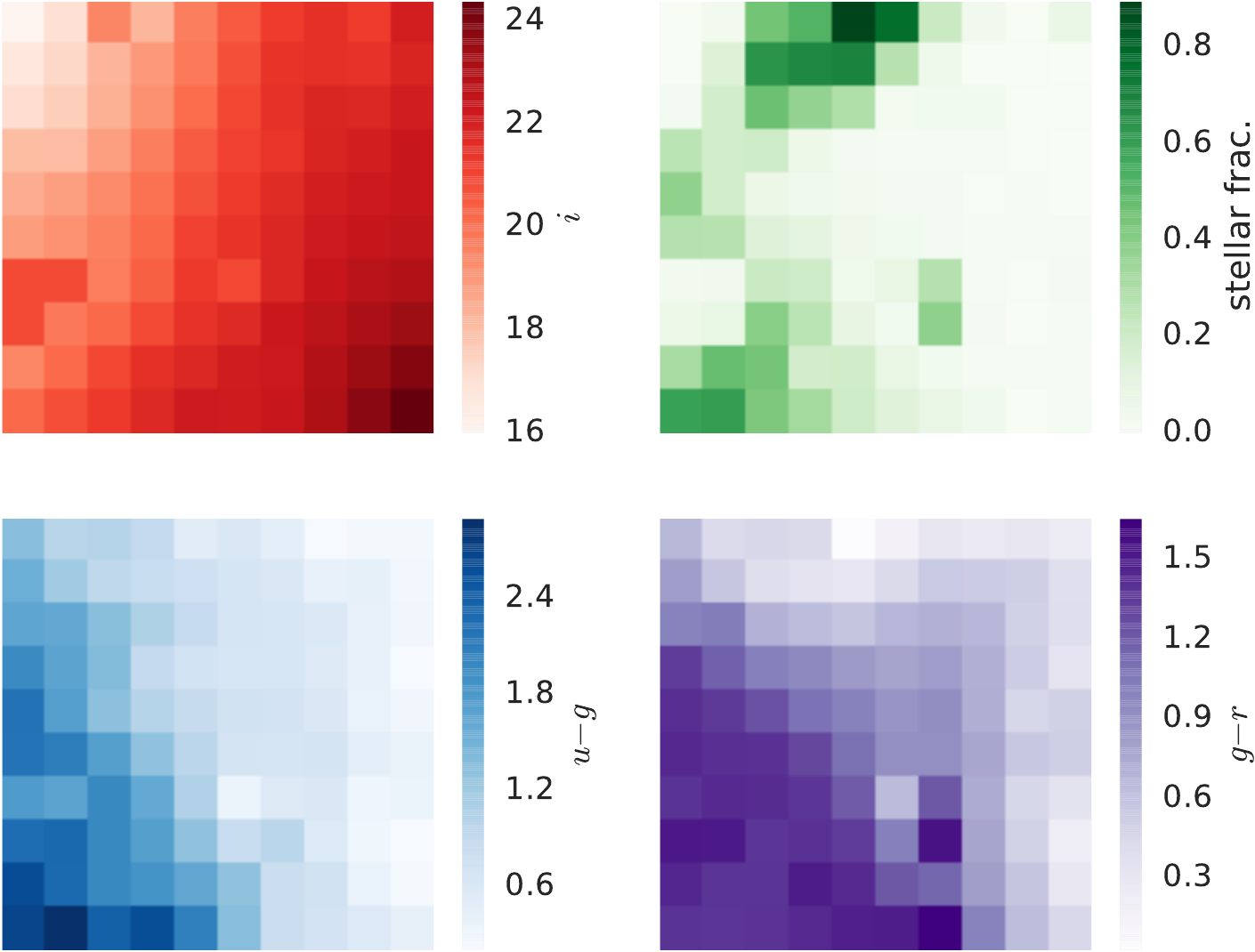}
  \caption{A two-dimensional 10$\times$10 SOM representation
           showing the mean $i$-band magnitude (top left),
           the fraction of true stars in each cell (top right),
           and the mean values of $u-g$ (bottom left) and $g-r$ (bottom right)
           for the cross-validation data.}
  \label{fig:som_colors}
\end{figure}

\begin{figure}
  \centering
  \includegraphics[width=\columnwidth]{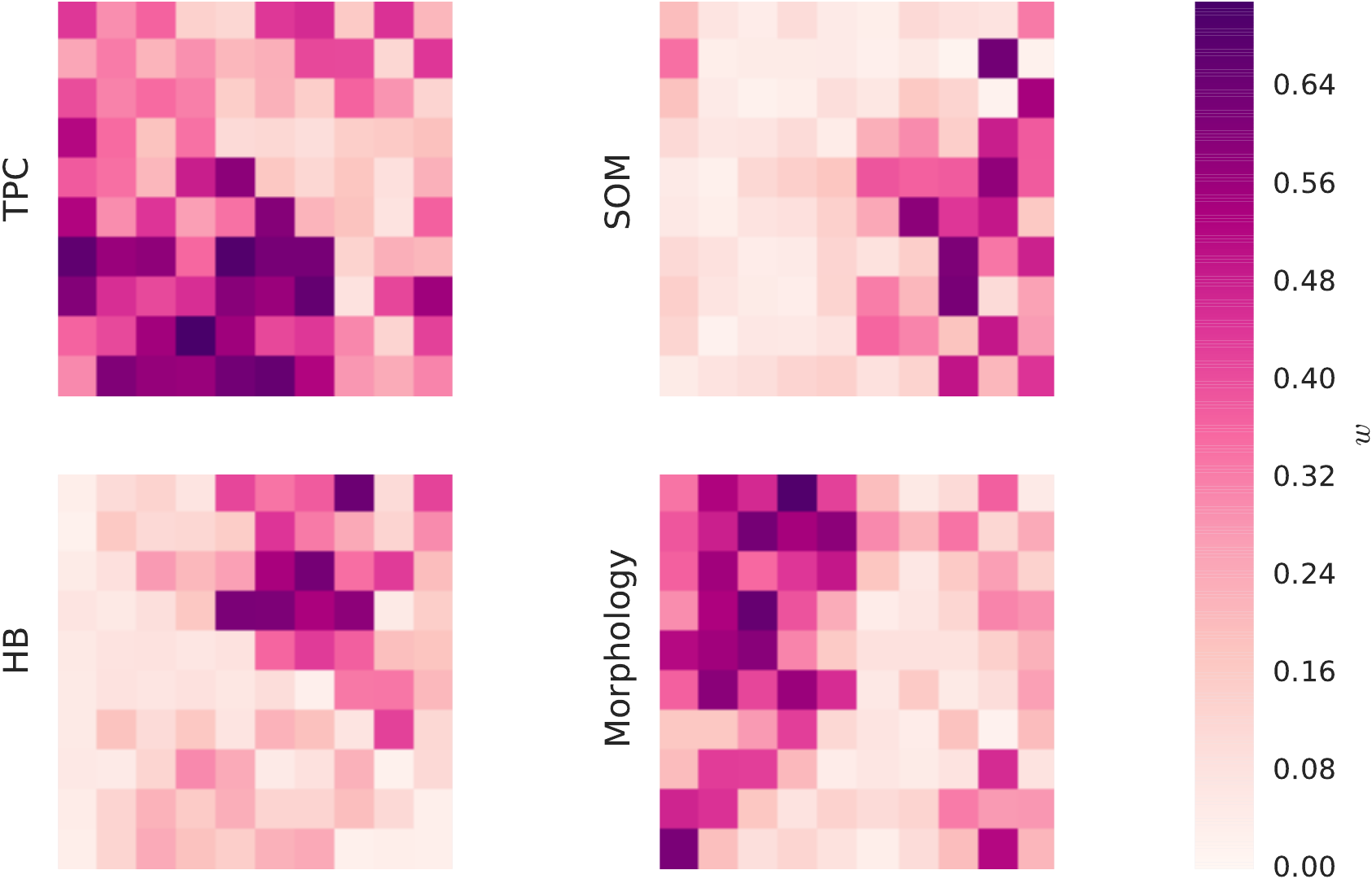}
  \caption{A two-dimensional 10$\times$10 SOM representation
           showing the relative weights for the BMC combination technique
           applied to the four individual methods for the CFHTLenS data.}
  \label{fig:weights}
\end{figure}

In Figure~\ref{fig:som_colors}, we show in the top left panel
the mean CFHTLenS $i$-band magnitude in each cell,
and in the top right panel the fraction of stars in each cell.
The bottom two panels show the mean $u-g$ and  $g-r$ colors in each cell.
These two-dimensional maps clearly show
the ability of the SOM to preserve relationships between sources
when it projects the full nine-dimensional space to the two-dimensional map.
We note that these SOM maps should only be used to provide guidance,
as the SOM mapping is a non-linear representation of all magnitudes and colors.

We can also use the same SOM from Figure~\ref{fig:som_colors}
to determine the relative weights for 
the four individual classification methods in each cell.
We present the four weight maps for the BMC technique 
in Figure~\ref{fig:weights}.
In these maps, a darker color indicates a higher weight,
or equivalently that the corresponding classifier
performs better in that region.
These weight maps demonstrate the variation in
the performance of the individual techniques across
the two-dimensional parameter space defined by the SOM.
Furthermore, since the maps in Figure~\ref{fig:som_colors}
and \ref{fig:weights} are constructed using the same SOM,
we can determine the region in the parameter space
where each individual technique performs better or worse.
Not surprisingly, the morphological separation
performs best in the top left corner of the weight map
in Figure~\ref{fig:weights},
which corresponds to the brightest CFHTLenS magnitudes $i \la 20$
in the $i$-band magnitude map of Figure~\ref{fig:som_colors}.
It is also clear that the SOM cells where the morphological
separation performs best have higher stellar fraction than
the other cells. 
On the other hand, TPC seems to perform best
in the region that corresponds to intermediate magnitudes
$20\la i \la22.5$ and $1.5 \la u-g \la 3.0$.
Our unsupervised learning method SOMc
performs relatively better at fainter magnitudes $i \ga 21.5$
with $0 \la u-g \la 0.5$ and $0 \la g-r \la 0.5$.
Although HB shows the worst performance
when there exists a high-quality training data set,
BMC still utilizes information from HB,
especially at intermediate magnitudes $20\la i \la22$.
Another interesting pattern is that
the four techniques seem complementary,
and they are weighted most strongly in different regions
of the SOM representation.

\begin{figure}
  \centering
  \includegraphics[width=\columnwidth]{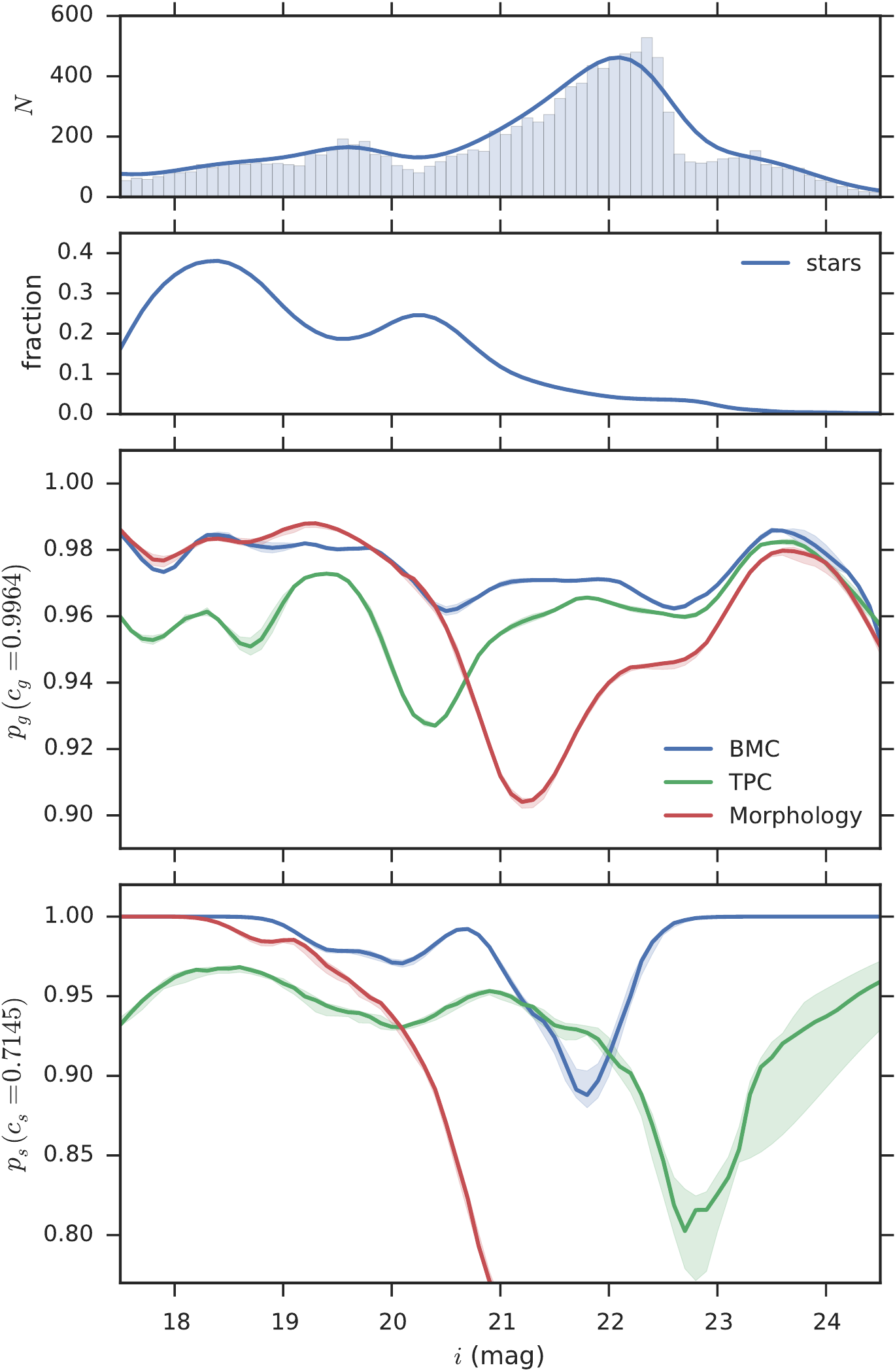}
  \caption{Purity as a function of the $i$-band magnitude
           as estimated by the kernel density estimation (KDE) method.
           The top panel shows the histogram with a bin size of 0.1 mag
           and the KDE for objects in the test set.
           The second panel shows the fraction of stars estimated by KDE
           as a function of magnitude.
           The bottom two panels compare
           the galaxy and star purity values for BMC, TPC, and 
           morphological separation as functions of magnitude.
           Results for BMC, TPC, and morphological separation are in
           blue, green, and red, respectively.
           The $1 \sigma$ confidence bands are estimated by
           bootstrap sampling.}
  \label{fig:purity_mag}
\end{figure}

In Figure~\ref{fig:purity_mag}, 
we compare the star and galaxy purity values
for BMC, TPC, and morphological separation
as functions of $i$-band magnitude.
We use the kernel density estimation \cite[KDE;][]{silverman1986density}
with the Gaussian kernel to smooth the fluctuations in the distribution.
Although morphological separation shows a slightly better performance
in galaxy purity at bright magnitudes $i \la 20$,
BMC outperforms both TPC and morphological separation
at faint magnitudes $i \ga 21$.
As the top panel shows,
the number count distribution peaks at $i \sim 22$,
and BMC therefore outperforms both TPC and morphological separation
for the majority of objects.
It is also clear that BMC outperforms TPC over all magnitudes.
BMC can presumably accomplish this by combining information from
all base classifiers,
\eg giving more weight to the morphological separation method
at bright magnitudes.
The bottom panel shows that
the star purity of morphological separation drops to $p_s < 0.8$
at fainter magnitudes $i > 21$.
This is expected, as our crude morphological separation classifies
every object as a galaxy beyond $i > 21$, and
purity measures the number of true stars classified as stars.
It is again clear that BMC outperforms both TPC and morphological separation
in star purity values over all magnitudes.

\begin{figure}
  \centering
  \includegraphics[width=\columnwidth]{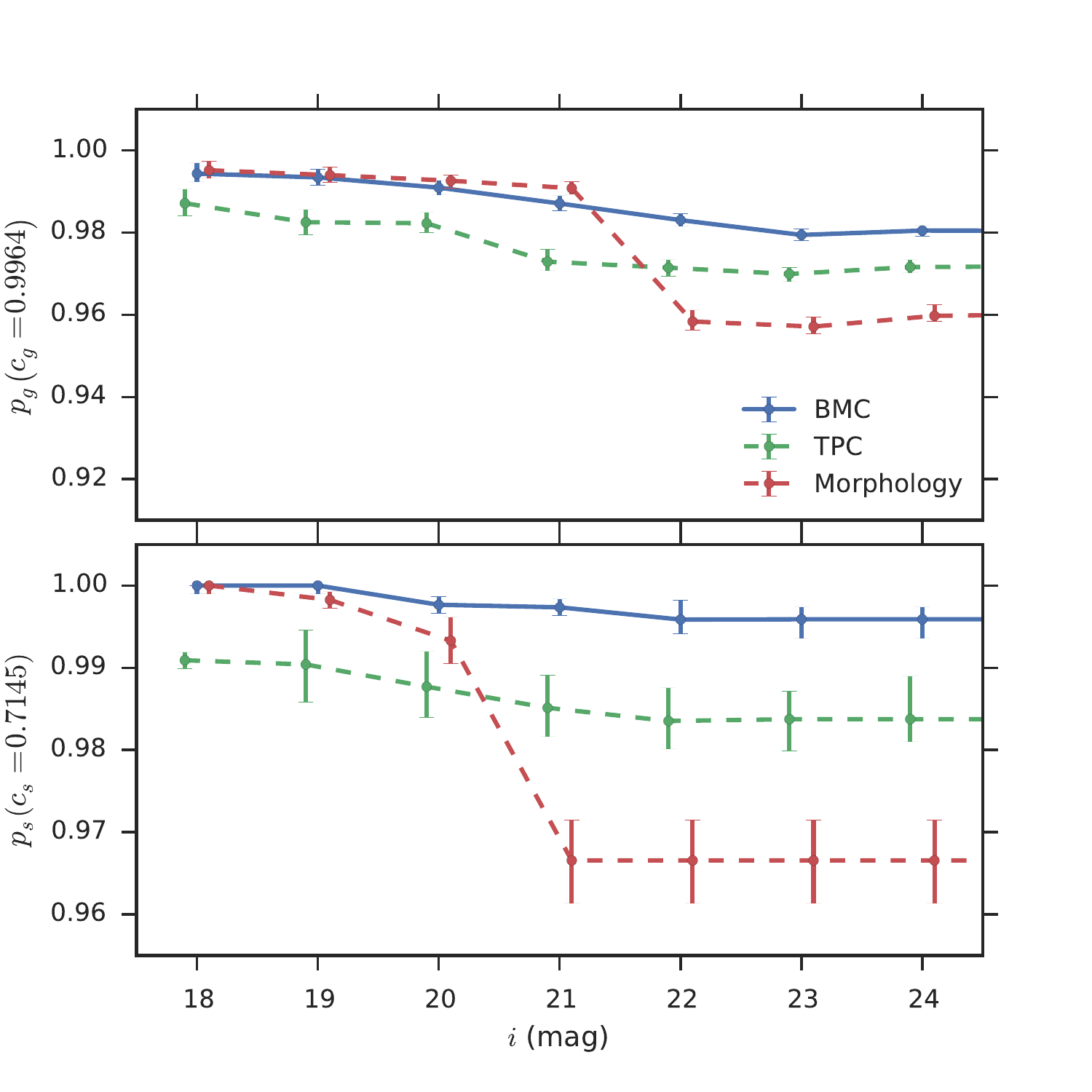}
  \caption{Cumulative purity as a function of the $i$-band magnitude.
           The upper panel compares
           the galaxy purity values for BMC (blue solid line),
           TPC (green dashed line), and 
           morphological separation (red dashed line).
           The lower panel compares the star purity.
           The $1 \sigma$ error bars are computed following the method
           of \citet{paterno2004calculating} to avoid the unphysical
           errors of binomial or Poisson statistics.}
  \label{fig:purity_mag_integrated}
\end{figure}

In Figure~\ref{fig:purity_mag_integrated}, we show the
cumulative galaxy and star purity values as functions of magnitude.
Although morphological separation performs better than TPC
at bright magnitudes, its purity values decrease as 
the magnitudes become fainter, and
TPC eventually outperforms morphological separation by 1--2\% at
$i > 21$.
BMC clearly outperforms both TPC and morphological separation,
and it maintains the overall galaxy purity of 0.980
up to $i \sim 24.5$.

\begin{figure}
  \centering
  \includegraphics[width=\columnwidth]{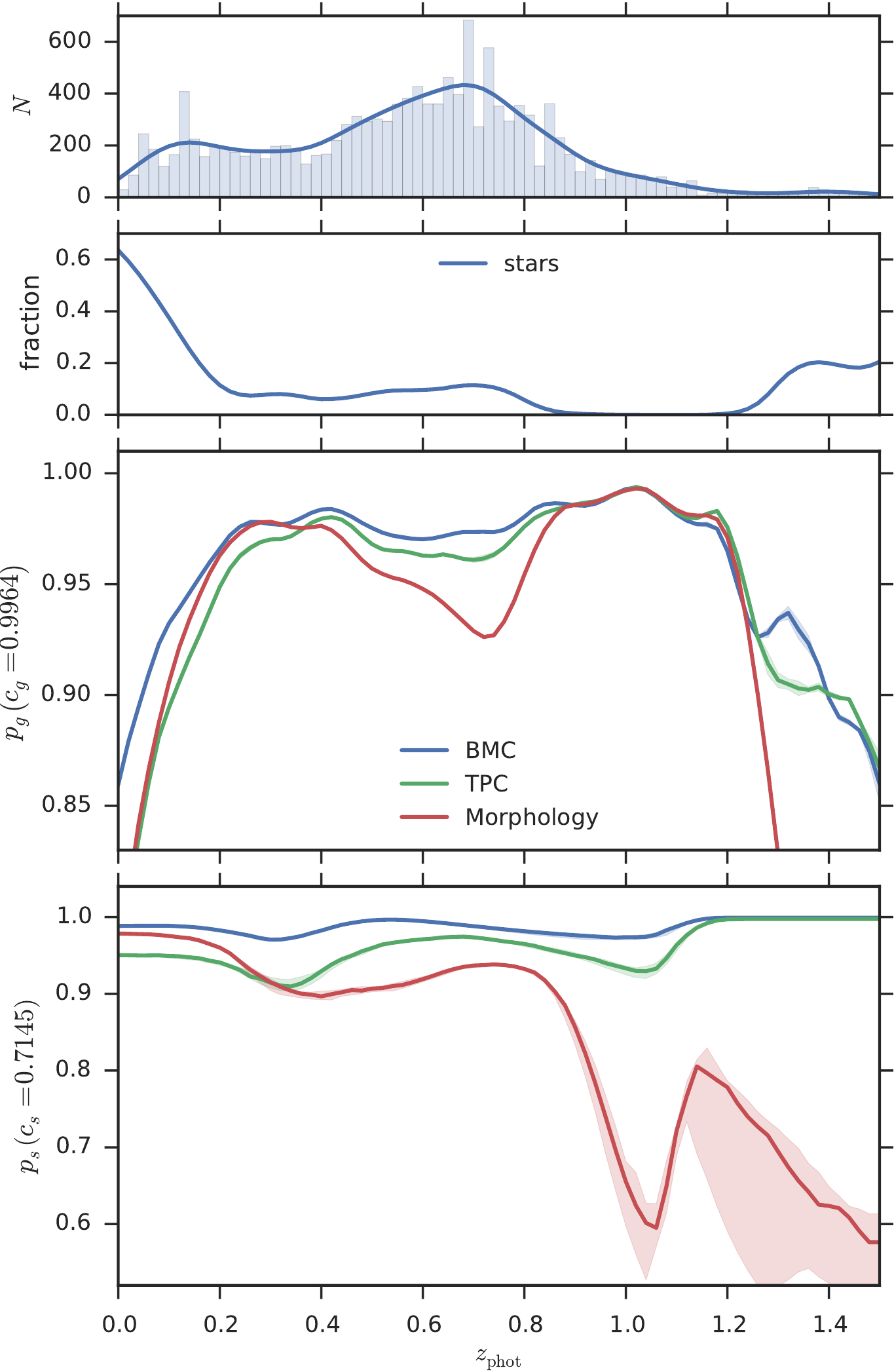}
  \caption{Similar to Figure~\ref{fig:purity_mag}
           but as a function of photo-$z$.
           The bin size of histogram in the top panel is 0.02.}
  \label{fig:purity_z}
\end{figure}

\begin{figure}
  \centering
  \includegraphics[width=\columnwidth]{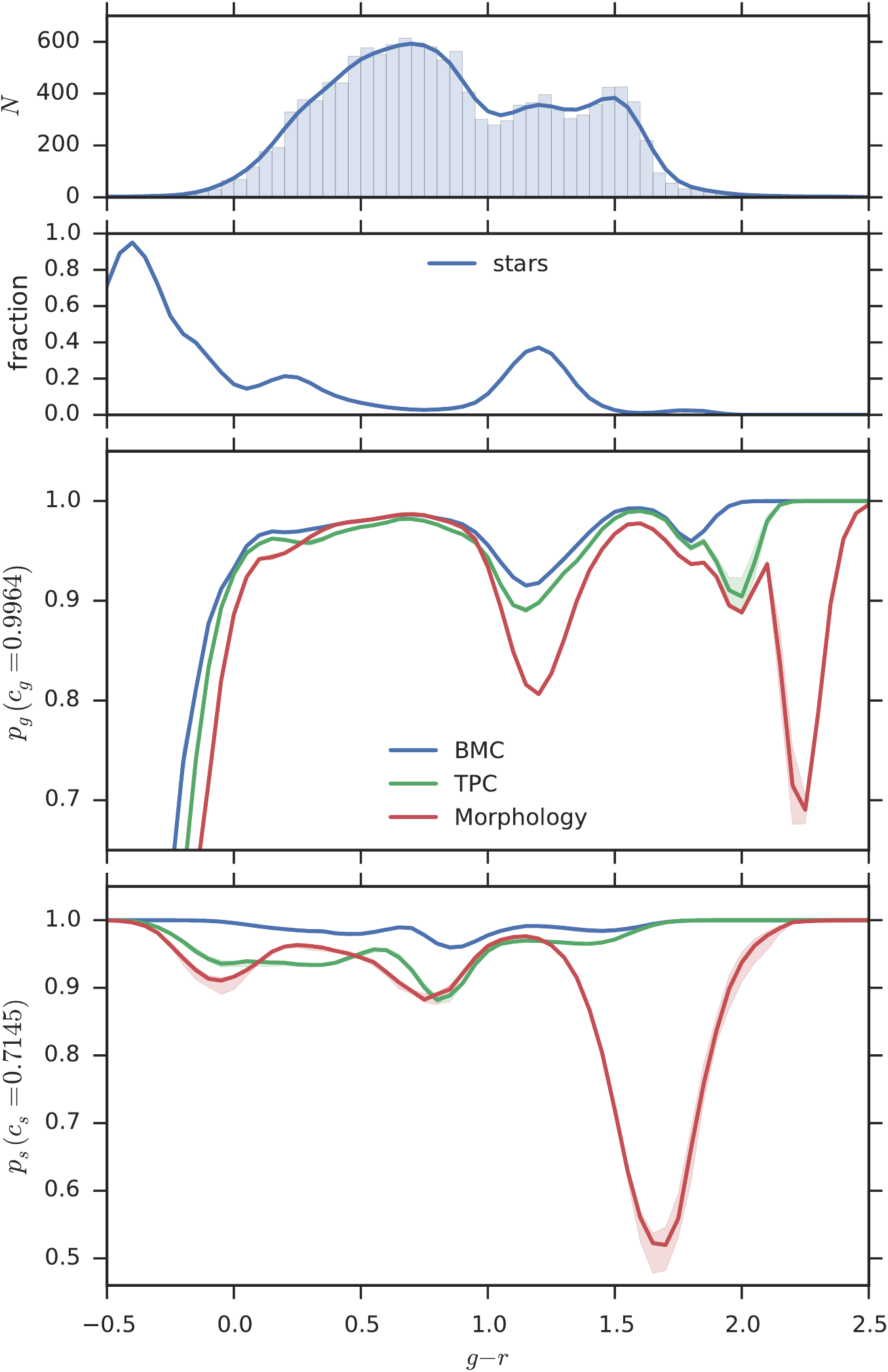}
  \caption{Similar to Figure~\ref{fig:purity_mag}
           but as a function of $g-r$ color.
           The bin size of histogram in the top panel is 0.05.}
  \label{fig:purity_g_r}
\end{figure}

We also show the star and galaxy purity values as functions of
photometric redshift estimate in Figure~\ref{fig:purity_z}.
Photo-$z$ is estimated with
the \textsc{BPZ} algorithm~\citep{benitez2000bayesian}
and provided with
the CFHTLenS photometric redshift catalogue~\citep{hildebrandt2012cfhtlens}.
The advantage of BMC over TPC and morphological separation is
now more pronounced in Figure~\ref{fig:purity_z}.
Although the morphological separation method outperforms BMC
at bright magnitudes in Figure~\ref{fig:purity_mag},
it is clear that BMC outperforms
both TPC and morphological separation over all redshifts.
We also present in Figure~\ref{fig:purity_g_r}
how the star and galaxy purity values vary
as a function of $g-r$ color.
It is again clear that BMC outperforms
both TPC and morphological separation over all $g-r$ colors.

\begin{figure}
  \includegraphics[width=\columnwidth]{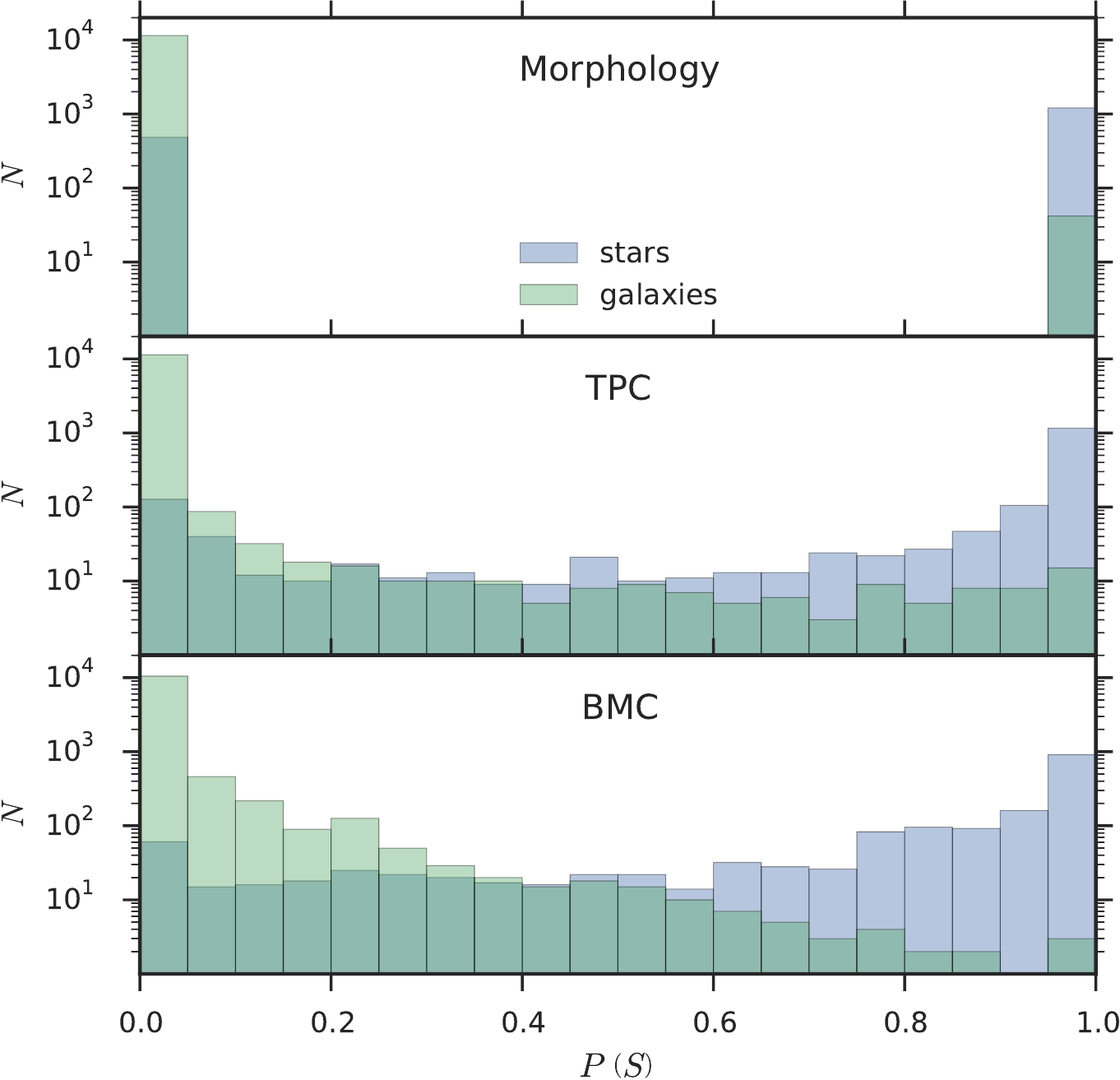}
  \caption{Histogram of the posterior probability that
           a source is a star for morphological separation (top),
           TPC (middle), and BMC (bottom)
           for a high-quality training data set.
           The true galaxies are in green, and true stars are in blue.
           The bin size is 0.05.}
  \label{fig:p_dist_all}
\end{figure}

In Figure~\ref{fig:p_dist_all},
we show the distribution of $P(S)$,
the posterior probability that an object is a star,
for BMC, TPC, and morphological separation.
It is interesting that the BMC technique assigns
a posterior star probability $P(S) \la 0.3$
to significantly more true galaxies than TPC,
and a probability $P(S) \ga 0.8$ to significantly fewer true galaxies.
By utilizing information from different types of classification techniques
in different parts of the parameter space,
BMC becomes more certain that an object is a star or a galaxy,
resulting in improvement of overall performance.

\subsection{Heterogeneous Training}
  \label{section:poor_training}
 
\begin{figure}
  \centering
  \includegraphics[width=\columnwidth]{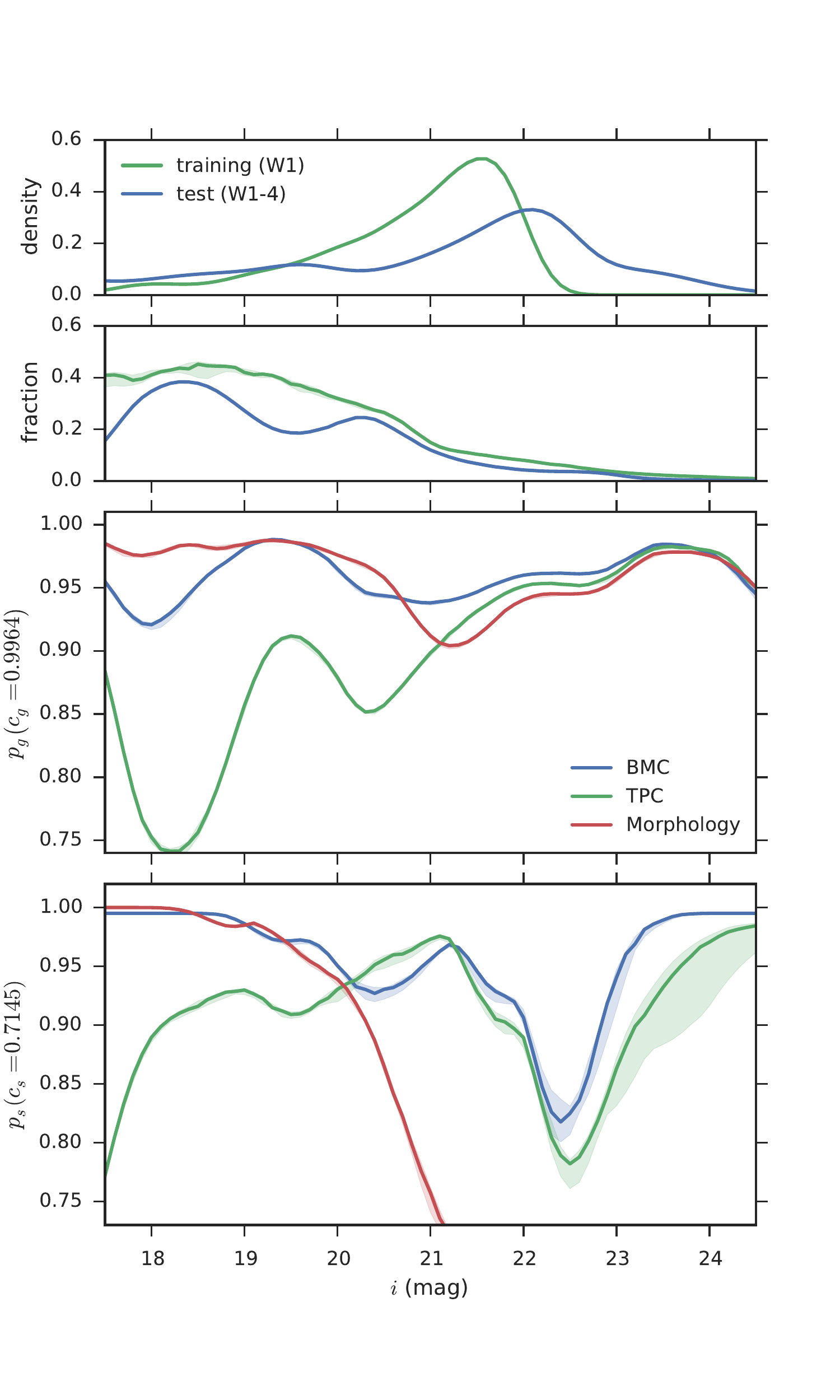}
  \caption{Purity as a function of the $i$-band magnitude
           for the reduced training data set.
           Top panel shows the histograms and KDEs
           for the number count distribution for
           the training (blue) and test (green) data set.
           The second panel shows the fraction of stars
           in the training and test data set in blue and green, respectively.
           The bottom two panels compare
           the galaxy and star purity values for BMC, TPC, 
           and morphological separation as functions of $i$-band magnitude.}
  \label{fig:purity_mag_cut}
\end{figure}

\begin{figure}
  \centering
  \includegraphics[width=\columnwidth]{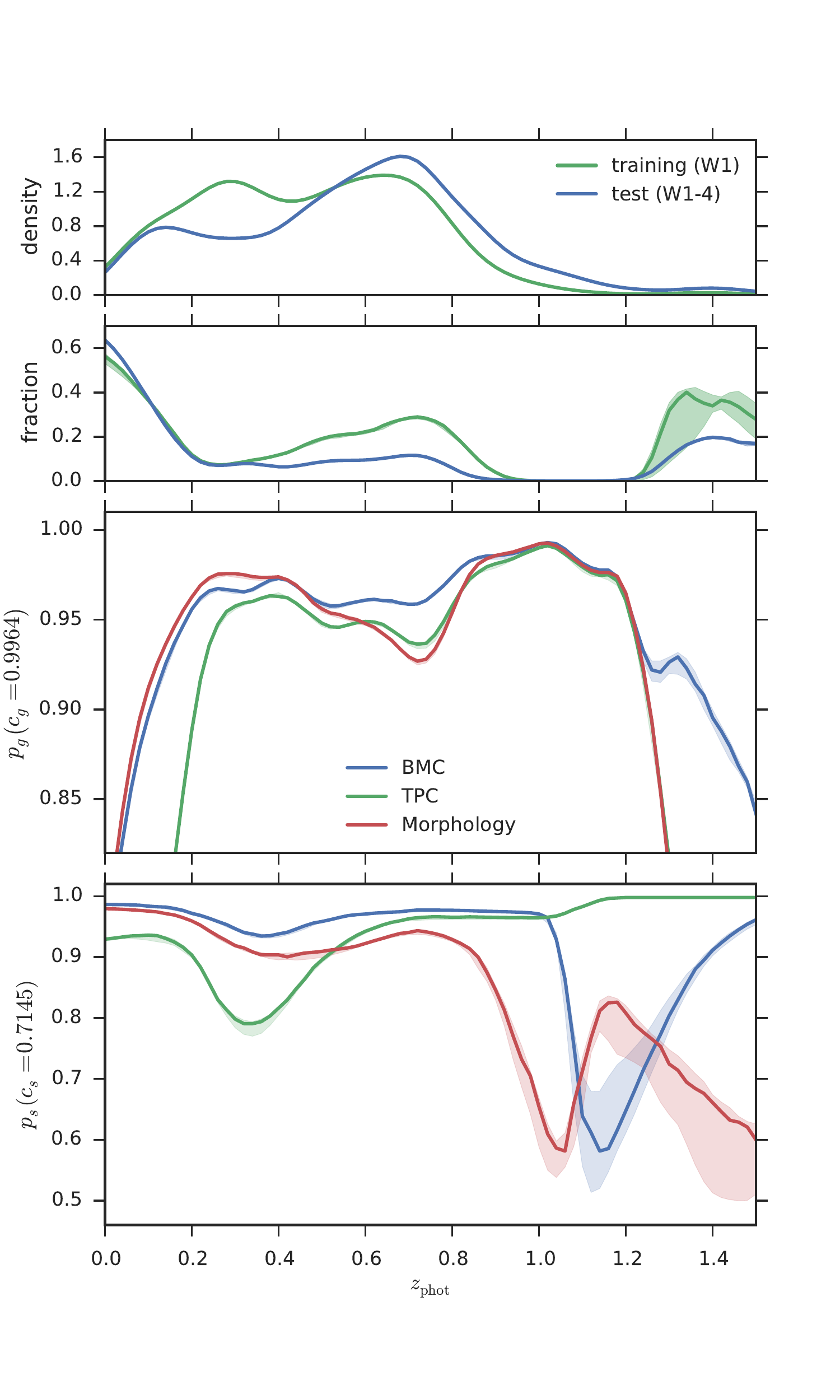}
  \caption{Similar to Figure~\ref{fig:purity_mag_cut}
           but as a function of photo-$z$.}
  \label{fig:purity_z_cut}
\end{figure}

\begin{figure}
  \centering
  \includegraphics[width=\columnwidth]{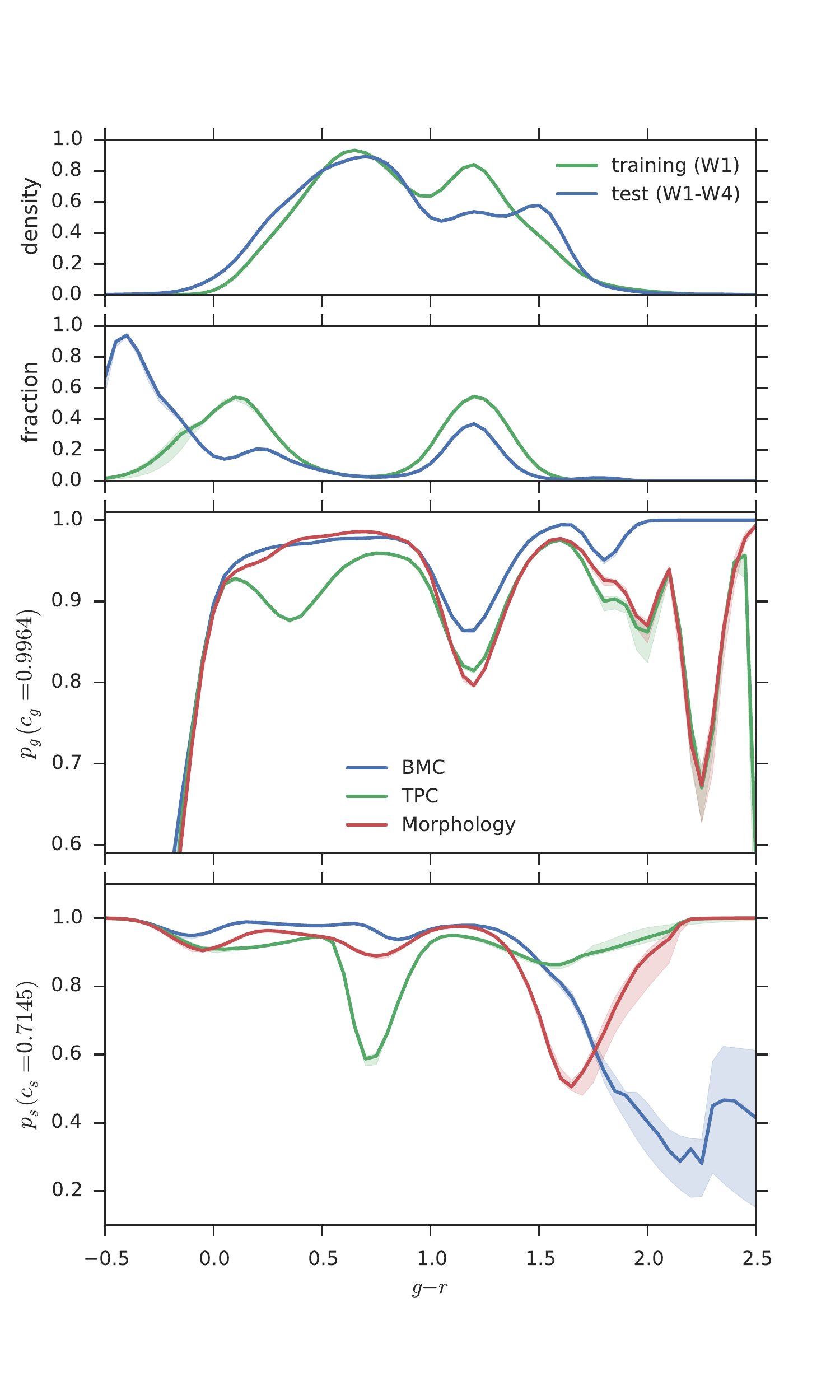}
  \caption{Similar to Figure~\ref{fig:purity_mag_cut}
           but as a function of $g-r$ color.}
  \label{fig:purity_g_r_cut}
\end{figure}

It is very costly in terms of telescope time to
obtain a large sample of spectroscopic observations
down to the limiting magnitude of a photometric sample.
Thus, we investigate the impact of training set quality
by considering a more realistic case
where the training data set is available
only for a small number of objects with bright magnitudes.
To emulate this scenario,
we only use objects that have spectroscopic labels
from the VVDS 0226-04 field (which is located within the CFHTLS W1 field)
and impose a magnitude cut of $i < 22.0$ in the training data,
leaving us a training set with only 1,365 objects.
We apply the same four star-galaxy classification techniques
and four combination methods,
and measure the performance of each technique on the same test data set
from Section~\ref{section:rich_training}.
As the top two panels of Figures~\ref{fig:purity_mag_cut},
\ref{fig:purity_z_cut}, and \ref{fig:purity_g_r_cut} show,
the demographics of objects in the training set
are different from the distribution of sources in the test set.
Thus, this also serves as a test of the efficacy of heterogeneous training.

\begin{table*}
  \caption{A summary of the classification performance metrics
           for the four individual methods
           and the four different classification combination methods
           when the training data set consists of
           only the sources that are in CFHTLS W1 field,
           has spectroscopic labels available from VVDS,
           and has $i < 22$.
           The definition of the metrics is summarized in
           Table~\ref{table:metrics}.
           The bold entries highlight the best performance values
           within each column.
           Note that some objects in the test set have bad or missing
           values (\eg $-99$ or $99$) in one or more attributes,
           which are included here (but are omitted, for example,
           in Figure~\ref{fig:purity_mag_cut_integrated} when the corrsponding
           attribute is not available.)}
  \centering
  \begin{tabular}{l c c c c c c}
  Classifier & AUC & MSE &
  $p_{g}\left(c_g=0.9964\right)$ & $p_{s}\left(c_s=0.7145\right)$ &
  $p_{g}\left(c_g=0.9600\right)$ & $p_{s}\left(c_s=0.2500\right)$ \\
  \hline
  TPC        & 0.9399 & 0.0511 & 0.9350 & 0.7060 & 0.9570 & 0.9747 \\
  SOMc       & 0.8861 & 0.0989 & 0.8843 & 0.4316 & 0.9165 & 0.6263 \\
  HB         & 0.9386 & 0.0760 & 0.9325 & 0.6911 & 0.9424 & 0.6918 \\
  Morphology & - & 0.0397 & 0.9597 & 0.9666 & - & - \\
  WA         & 0.9600 & 0.0536 & 0.9208 & 0.8818 & 0.9757 & 0.9815 \\
  BoM        & 0.9587 & 0.1511 & 0.9658 & 0.9862 & 0.9790 & 0.9977 \\
  Stacking   & 0.9442 & 0.1847 & 0.9561 & 0.9309 & 0.9664 & 0.9983 \\
  BMC        & \textbf{0.9738} & \textbf{0.0291} & \textbf{0.9696} & $\textbf{0.9862}$ & $\textbf{0.9856}$ & \textbf{1.0000} \\
\end{tabular}
  \label{table:metrics_cut}
\end{table*}

We present in Table~\ref{table:metrics_cut}
the same six metrics for each method,
and highlight the best method for each metric.
Overall, the results obtained for the reduced data set are remarkable.
With a smaller training set, our training based methods, TPC and SOMc,
suffer a significant decrease in performance.
The performance of morphological separation and HB
is essentially unchanged from Table~\ref{table:metrics_all}
as they do not depend on the training data.
Without sufficient training data,
the advantage of combining the predictions of different classifiers
is more obvious.
Even WA, the simplest of combination techniques, outperforms
all individual classification techniques in four metrics,
AUC, $p_s$ at $c_s=0.7145$, $p_g$ at $c_g=0.9600$, and $p_s$ at $c_s=0.2500$.
Although BoM always chooses TPC as the best model
when we have a high-quality training set,
it now chooses various methods in different bins
and outperforms all base classifiers.
While the performance of the stacking technique is only slightly worse
than that of BMC when we have a high-quality training set,
stacking now fails to outperform morphological separation.
BMC shows an impressive performance and 
outperforms all other classification techniques in all six metrics.
Overall, the improvements are small but still significant
since these metrics are averaged over the full test data.

\begin{figure}
  \includegraphics[width=\columnwidth]{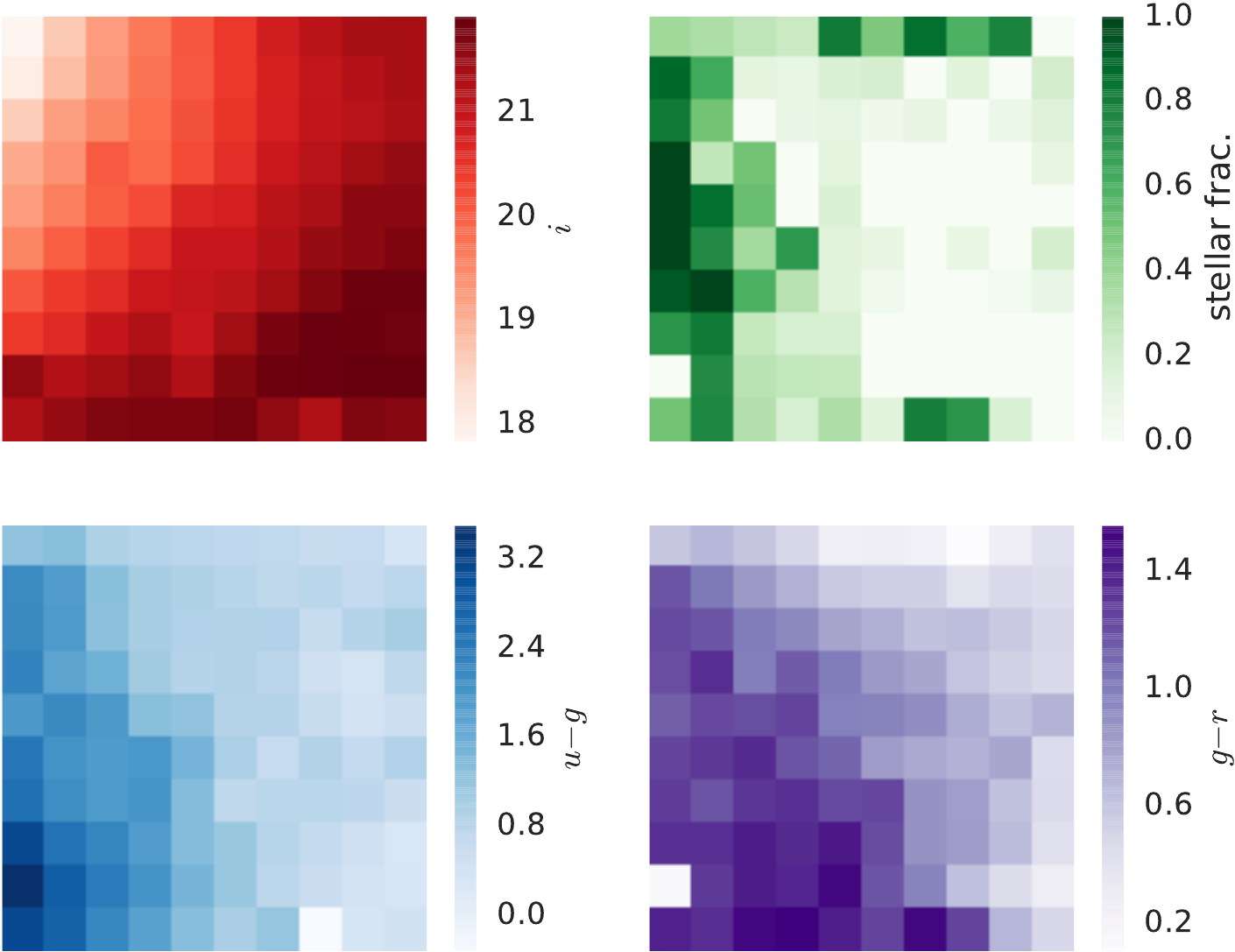}
  \caption{Similar to Figure~\ref{fig:som_colors}
           but for the reduced training data set. }
  \label{fig:som_colors_cut}
\end{figure}

\begin{figure}
  \includegraphics[width=\columnwidth]{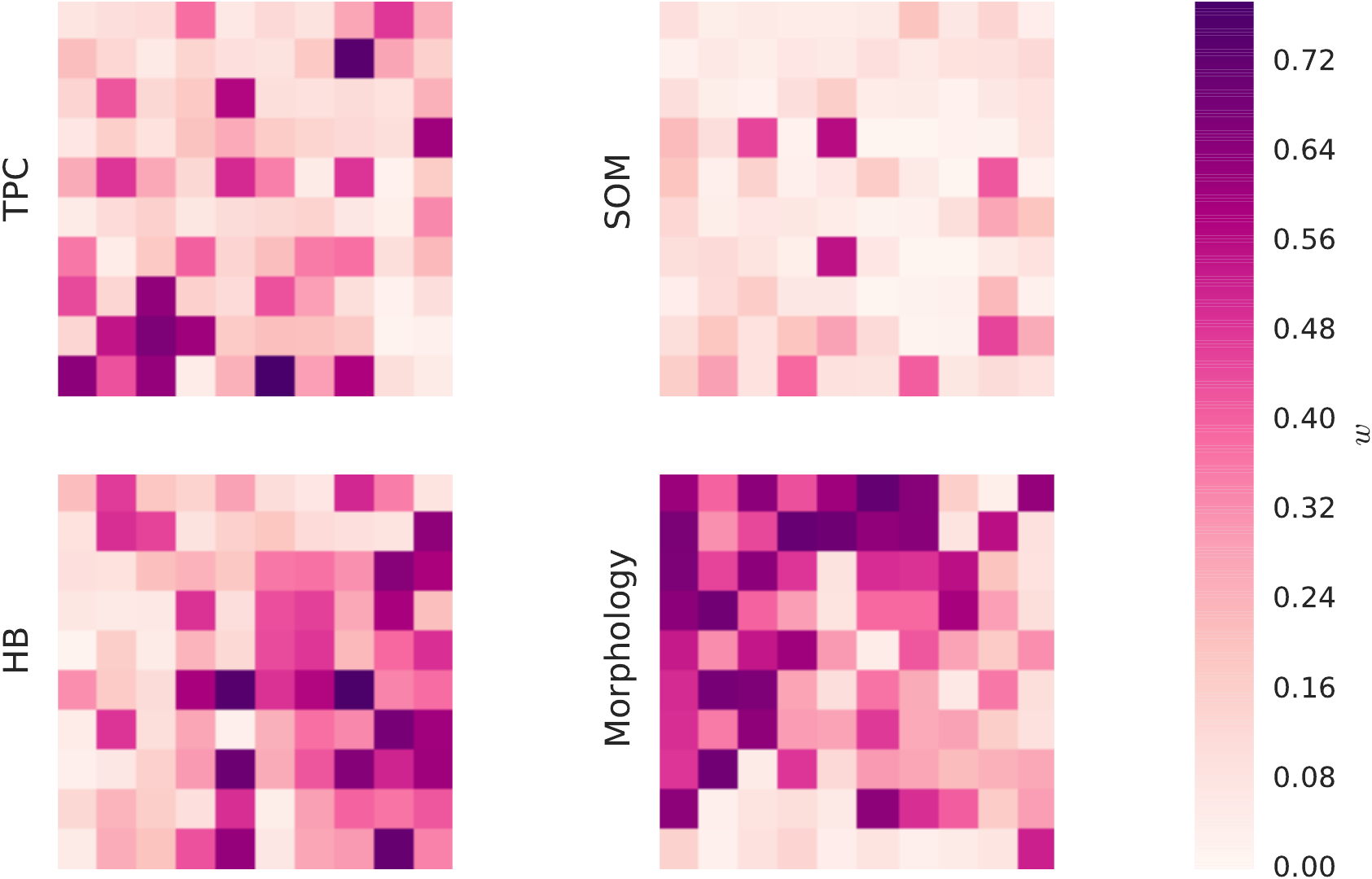}
  \caption{Similar to Figure~\ref{fig:weights}
           but for the reduced training data set. }
  \label{fig:weights_cut}
\end{figure}

In Figure~\ref{fig:weights_cut}, we again show the $10\times10$
two-dimensional weight map defined by the SOM.
When the quality of training data is relatively poor,
the performance of training based algorithms will decrease,
while the performance of template fitting algorithms
or morphological separation methods
is independent of training data.
Thus, when the weight maps of Figure~\ref{fig:weights}
and Figure~\ref{fig:weights_cut} are visually compared,
it is clear that
the BMC algorithm now uses more information from
morphological separation and HB,
while it uses considerably less information from 
our training based algorithms, TPC and SOMc.
Not surprisingly, the morphological separation method
performs best at bright magnitudes,
and BMC assigns more weight to HB at fainter magnitudes.

We present the star and galaxy purity values as functions of
$i$-band magnitude in Figure~\ref{fig:purity_mag_cut}.
The normalized density distribution as a function of magnitude
in the top panel
and the stellar distribution in the second panel
clearly show that the demographics of the training set and 
that of the test set are different.
Since the training set is cut at $i < 22$,
the density distribution falls off sharply around $i \sim 22$
and has a higher fraction of stars than the test set.
Compared to the purity values in Figure~\ref{fig:purity_mag},
TPC now suffers a significant decrease in star and galaxy purity.
However, the purity of BMC does not show such a significant drop
and decreases by only 2--5\%.
As suggested by the weight maps in Figure~\ref{fig:weights_cut},
BMC can accomplish this by shifting the relative weights assigned to
each base classifier in different SOM cells.
As the quality of training set worsens,
BMC assigns less weight to training based methods
and more weight to HB and morphological separation.

\begin{figure}
  \centering
  \includegraphics[width=\columnwidth]{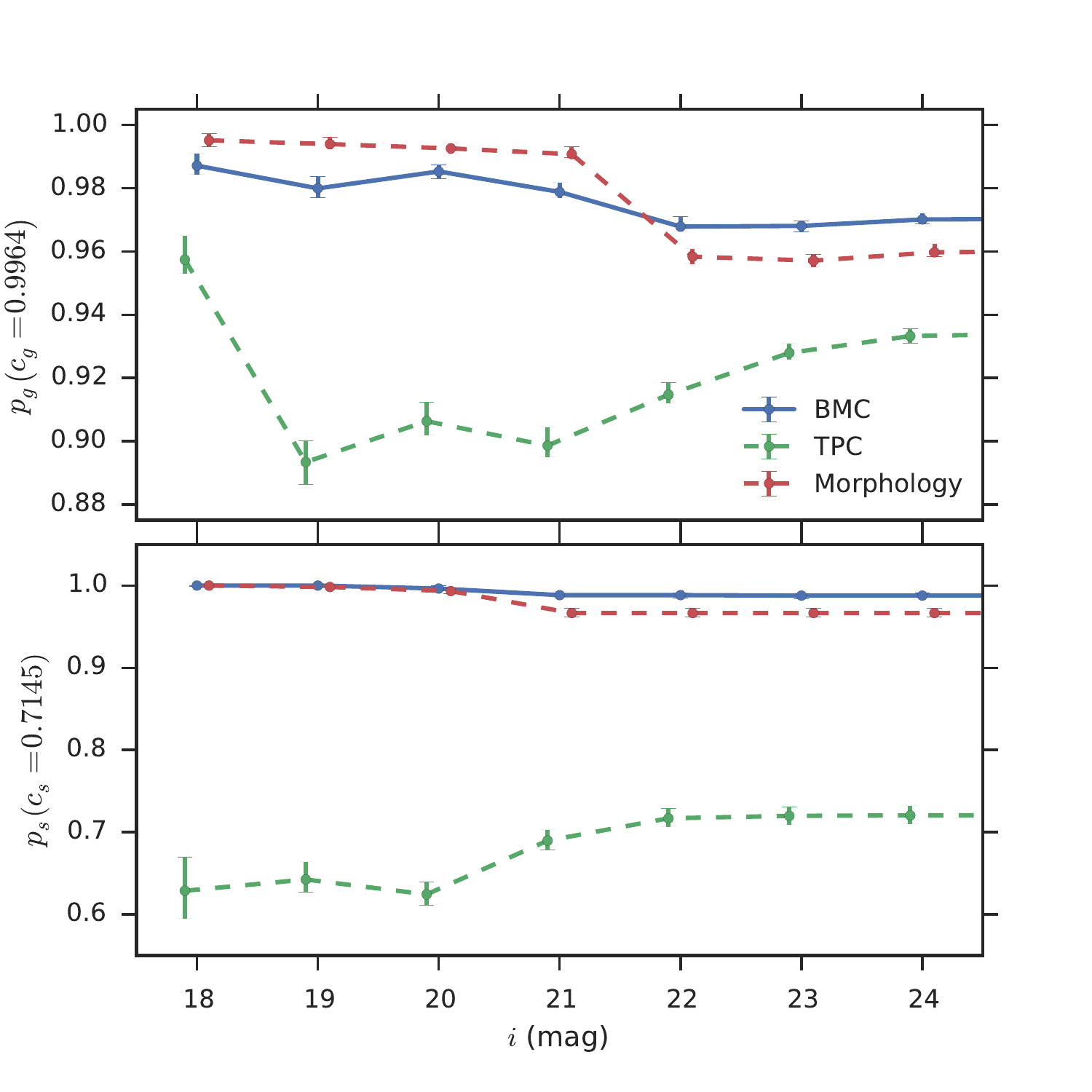}
  \caption{Similar to Figure~\ref{fig:purity_mag_integrated}
           but for the reduced training data set.}
  \label{fig:purity_mag_cut_integrated}
\end{figure}

In Figure~\ref{fig:purity_mag_cut_integrated}, we show the
cumulative galaxy and star purity values as functions of magnitude.
Compared to Figure~\ref{fig:purity_mag_integrated},
the drop in the performance of TPC is clear.
However, even when some classifiers have been trained on
a significantly reduced training set,
BMC maintains a galaxy purity of 0.970 and a star purity of 1.0
up to $i \sim 24.5$, and it sill outperforms morphological separation
at fainter magnitudes $i \ga 21$.

We also show the star and galaxy purity values as functions of
photo-$z$ in Figure~\ref{fig:purity_z_cut}
and as functions of $g-r$ in Figure~\ref{fig:purity_g_r_cut}.
Compared to Figure~\ref{fig:purity_z} and \ref{fig:purity_g_r},
the performance of BMC becomes worse in some photo-$z$ and $g-r$ bins.
However, this drop in performance seems to be confined to
only a small number of objects in particular regions of
the parameter space,
and BMC still outperforms both TPC and morphological separation
for the majority of objects.

\begin{figure}
    \centering
  \includegraphics[width=\columnwidth]{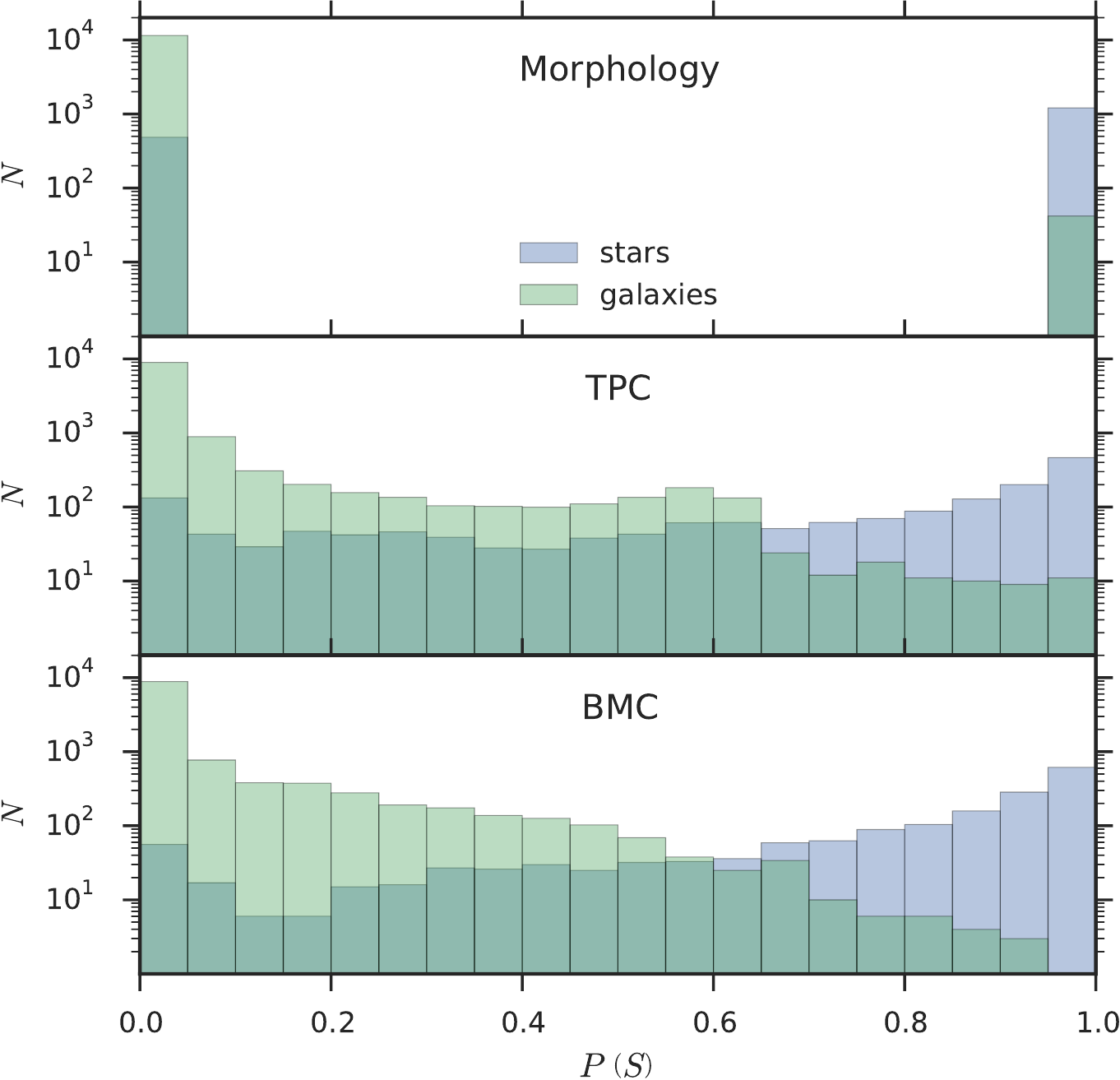}
  \caption{Similar to Figure~\ref{fig:p_dist_all}
           but for the reduced training data set.}
  \label{fig:p_dist_cut}
\end{figure}

Compared to Figure~\ref{fig:p_dist_all},
the difference between the posterior star probability distribution of
TPC and that of BMC is now more pronounced in Figure~\ref{fig:p_dist_cut}.
The $P\left(S\right)$ distribution of BMC for true galaxies
falls off sharply at $P\left(S\right)\approx0.95$,
and BMC does not assign a star probability
$P(S) \ga 0.95$ to any true galaxies,
On the other hand, both TPC and morphological separation 
classify some true galaxies as stars with absolute certainty.

\subsection{The Quality of Training Data}
  \label{section:quality_training}

The combination techniques that we have demonstrated so far use
two training based algorithms as base classifiers.
Ideally, the training data should mirror the entire parameter space
occupied by the data to be classified.
Yet we have seen in Section~\ref{section:poor_training}
that the BMC technique does reliably
extrapolate past the limits of the training data,
even when some base classifiers are trained on a low-quality training data set.
In this section, we further investigate
if and where BMC begins to break down
by imposing various magnitude, photo-$z$, and color cuts
to change the size and composition of the training set.

\begin{figure}
    \centering
    \includegraphics[width=\columnwidth]{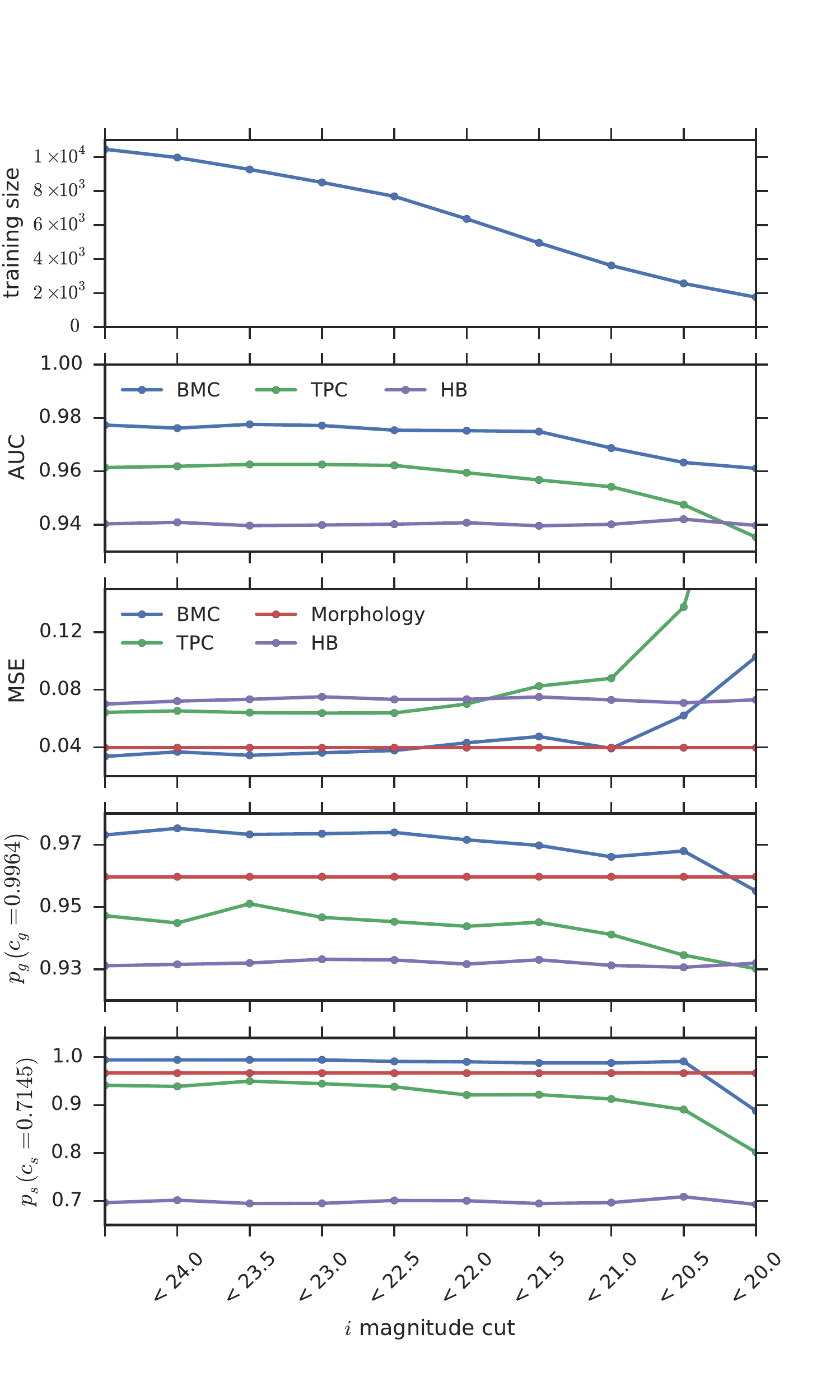}
    \caption{The classification performance metrics for
      BMC (blue), TPC (green), morphology (red), and HB (purple)
      as applied to the CFHTLenS data in the VVDS field
      with various magnitude cuts.
      The top panel shows the number of sources in the training set
      at corresponding magnitude cuts.
      We show only one of the four combination methods, BMC,
      which has the best overall performance.}
    \label{fig:perform_mag_cut}
\end{figure}

In Figure~\ref{fig:perform_mag_cut}, we present
a visual comparison between different classification techniques,
when various magnitude cuts are applied on the training data,
and the performance is measured on the same test set
from Section~\ref{section:rich_training} and \ref{section:poor_training}.
It is not surprising that the performance of TPC decreases
as we decrease the size of training set
by imposing more restrictive magnitude cuts,
while the performance of HB and morphological separation
is essentially unchanged.
However, the effect of change in size and composition of the training set
is significantly mitigated by the use of the BMC technique.
BMC outperforms both HB and TPC in all four metrics,
even when the training set is restricted to $i < 20.0$.
BMC also consistently outperforms morphological separation
until we impose a magnitude cut of $i < 20.0$ on the training data,
beyond which point BMC finally performs worse than morphological separation.
It is remarkable that BMC is able to reliably extrapolate
past the training data to $i \sim 24.5$,
the limiting magnitude of the test set, and outperform HB, TPC,
and morphological separation in all performance metrics,
even the demographics of training set do not accurately sample 
the data to be classified.

\begin{figure}
    \centering
    \includegraphics[width=\columnwidth]{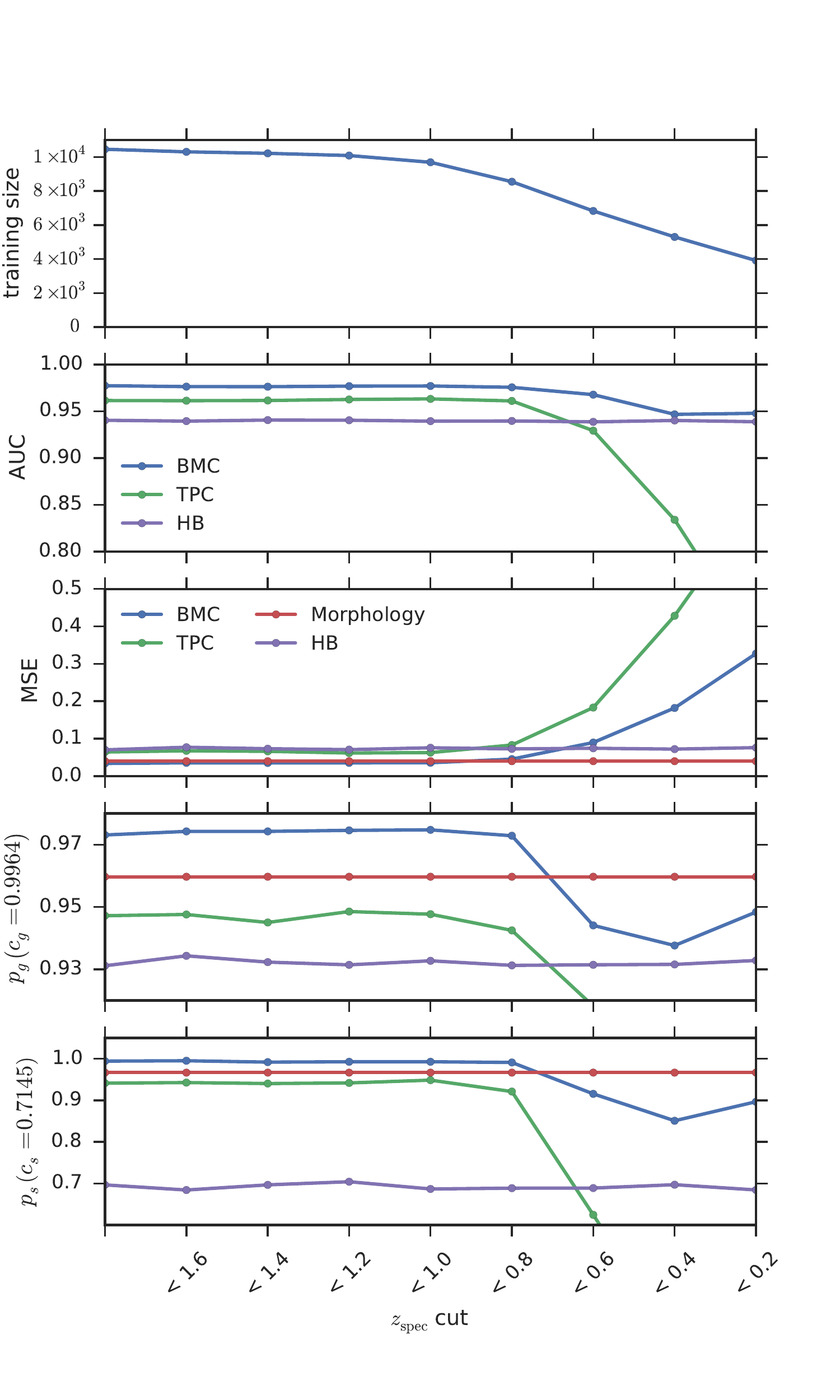}
    \caption{Similar to Figure~\ref{fig:perform_mag_cut}
        but using $z_{\text{spec}}$ cuts.}
    \label{fig:perform_z_cut}
\end{figure}

Similarly, we impose various spectroscopic redshift cuts 
on the training data in Figure~\ref{fig:perform_z_cut}.
Since all stars have $z_{\rmn{spec}}$ values
close to zero, we are effectively changing the demographics
of training set by keeping all stars and gradually removing
galaxies with high redshifts.
BMC begins to perform worse than morphological separation
when a conservative cut of $z_{\rmn{spec}} < 0.6$ is imposed.
However, it is again clear that BMC is able to
utilize information from HB and morphological separation to
mitigate the drop in the performance of TPC.

\begin{figure}
    \centering
    \includegraphics[width=\columnwidth]{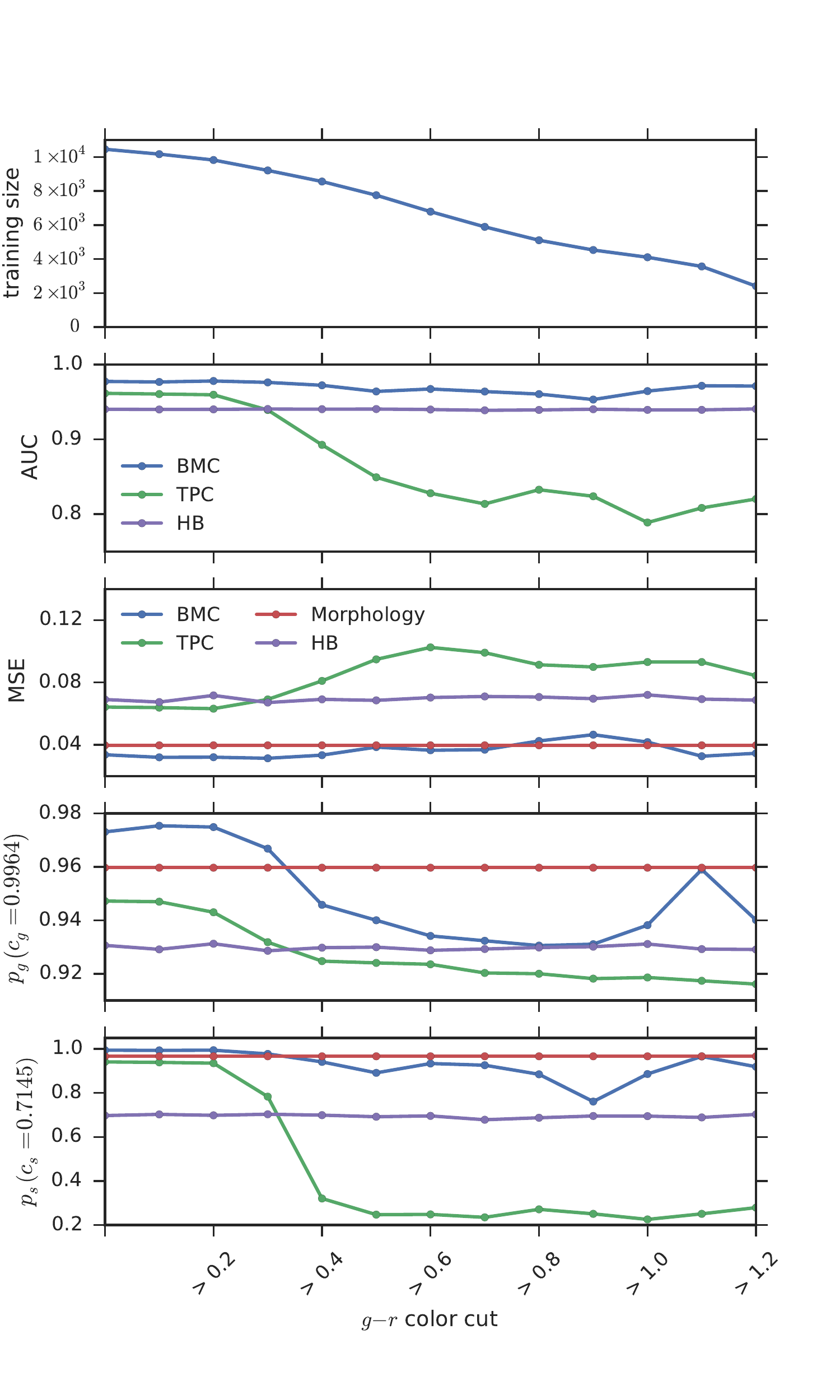}
    \caption{Similar to Figure~\ref{fig:perform_mag_cut}
        but using $g - r$ color cuts.}
    \label{fig:perform_g_r_cut}
\end{figure}

In Figure~\ref{fig:perform_g_r_cut},
we decrease the size of training set
by keeping red objects and gradually removing blue objects.
A color cut seems to have a more pronounced effect
on the performance of TPC and BMC,
which perform worse than morphological separation
when the training set is restricted to $g - r > 0.4$.
The performance depends more strongly on the color distribution, because 
a significant fraction of blue objects consists of stars,
while objects with fainter magnitudes and higher redshifts
are mostly galaxies.
We can verify this in Figure~\ref{fig:som_colors},
where the darker (higher stellar fraction) cells in the upper middle region
of the stellar fraction map (top right panel)
have bright magnitudes $i \la 20$
in the $i$-band magnitude map (top left panel)
and blue colors $g-r \la 0.5$ in the $g-r$ color map (bottom right panel).
On the other hand, the darker (fainter magnitude) cells
in the right-hand side of the $i$-band magnitude map
have almost no stars in them and
are represented by bright (low stellar fraction)
cells in the stellar fraction map.
Thus, these results indicate that the performance of training based
methods depends more strongly on the composition of training data
than on the size, and it is necessary to have
a sufficient number of the minority class
in the training data set to ensure optimal performance.

\section{Conclusions}
  \label{section:conclusions}

We have presented and analyzed a novel star-galaxy classification framework
for combining star-galaxy classifiers using the CFHTLenS data.
In particular, we use four independent classification techniques:
a morphological separation method;
TPC, a supervised machine learning technique
based on prediction trees and a random forest;
SOMc, an unsupervised machine learning approach
based on self-organizing maps and a random atlas;
and HB, a Hierarchical Bayesian template-fitting method
that we have modified and parallelized.
Both TPC and SOMc algorithms are currently available within
a software package named
\textsc{MLZ}\footnote{http://lcdm.astro.illinois.edu/code/mlz.html}.
Our implementation of HB and BMC,
as well as \textsc{IPython} notebooks that have been used to
produce the results in this paper,
are available at \url{https://github.com/EdwardJKim/astroclass}.  

Given the variety of star-galaxy classification methods we are using,
we fully expect the relative performance
of the individual techniques to vary across
the parameter space spanned by the data.
We therefore adopt the binning strategy, where
we allow different classifier combinations 
in different parts of parameter space
by creating two-dimensional self-organizing maps of
the full multi-dimensional magnitude-color space.
We apply different star-galaxy classification techniques
within each cell of this map,
and find that the four techniques are weighted most strongly
in different regions of the map.   

Using data from the CFHTLenS survey,
we have considered different scenarios:
when an excellent training set is available with spectroscopic labels from
DEEP2, SDSS, VIPERS, and VVDS, and
when the demographics of sources in a low-quality training set
do not match the demographics of objects in the test data set.
We demonstrate that the Bayesian Model Combination (BMC) technique improves
the overall performance over any individual classification method
in both cases.
We note that \citet{carrascokind2014exhausting} analyzed
different techniques for combining
photometric redshift probability density functions (photo-$z$ PDFs)
and also found that BMC is in general the best
photo-$z$ PDF combination technique.   

The problem of star-galaxy classification is a rich area
for future research.
It is unclear if sufficient training data will be available in future
ground-based surveys. Furthermore, in large sky surveys such as DES and LSST,
photometric quality is not uniform across the sky,
and a purely morphological classifier alone will not be sufficient,
especially at faint magnitudes.
Given the efficacy of our approach,
classifier combination strategies are likely the optimal approach
for currently ongoing and forthcoming photometric surveys.
We therefore plan to apply the combination technique described in this paper
to other surveys such as the DES.
Our approach can also be extended more broadly to
classify objects that are neither stars nor galaxies (\eg quasars).
Finally, future studies could explore the use of multi-epoch data,
which would be particularly useful for the next generation of
synoptic surveys.

\section*{Acknowledgements}

The authors thank the referee for a careful reading of the manuscript
and comments that improved this work.
We thank Ignacio Sevilla for helpful and insightful conversations.
We acknowledge support from the 
National Science Foundation Grant No.\ AST-1313415.
RJB acknowledges support as an Associate
within the Center for Advanced Study at the University of Illinois.

This work used the Extreme Science and Engineering Discovery Environment
(XSEDE), which is supported by National Science Foundation grant number
ACI-1053575.

This work is based on observations obtained with MegaPrime/MegaCam, a
joint project of CFHT and CEA/DAPNIA, at the Canada-France-Hawaii
Telescope (CFHT) which is operated by the National Research Council
(NRC) of Canada, the Institut National des Sciences de l'Univers of
the Centre National de la Recherche Scientifique (CNRS) of France, and
the University of Hawaii. This research used the facilities of the
Canadian Astronomy Data Centre operated by the National Research
Council of Canada with the support of the Canadian Space Agency.
CFHTLenS data processing was made possible thanks to significant
computing support from the NSERC Research Tools and Instruments grant
program.

Funding for the DEEP2 survey has been provided by NSF grants AST-0071048,
AST-0071198, AST-0507428, and AST-0507483. 

Funding for SDSS-III has been provided by the Alfred P. Sloan Foundation, the
Participating Institutions, the National Science Foundation, and the U.S.
Department of Energy Office of Science. The SDSS-III web site is
http://www.sdss3.org/.

SDSS-III is managed by the Astrophysical Research Consortium for the
Participating Institutions of the SDSS-III Collaboration including the
University of Arizona, the Brazilian Participation Group, Brookhaven National
Laboratory, Carnegie Mellon University, University of Florida, the French
Participation Group, the German Participation Group, Harvard University, the
Instituto de Astrofisica de Canarias, the Michigan State/Notre Dame/JINA
Participation Group, Johns Hopkins University, Lawrence Berkeley National
Laboratory, Max Planck Institute for Astrophysics, Max Planck Institute for
Extraterrestrial Physics, New Mexico State University, New York University,
Ohio State University, Pennsylvania State University, University of Portsmouth,
Princeton University, the Spanish Participation Group, University of Tokyo,
University of Utah, Vanderbilt University, University of Virginia, University
of Washington, and Yale University.

This paper uses data from the 
VIMOS Public Extragalactic Redshift Survey (VIPERS).
VIPERS has been performed using the ESO Very Large Telescope, under the "Large
Programme" 182.A-0886. The participating institutions and funding agencies are
listed at http://vipers.inaf.it/.

This research uses data from the VIMOS VLT Deep Survey, obtained from the VVDS
database operated by Cesam, Laboratoire d'Astrophysique de Marseille, France.

\footnotesize{
\bibliographystyle{mn2e}
\bibliography{sg_paper}

\begin{thebibliography}{}

\bibitem[\protect\citeauthoryear{{Ahn} et~al.,}{{Ahn}  et~al.}{2014}]{Ahn2014}
{Ahn} C.~P.,  et~al., 2014, \apjs, 211, 17

\bibitem[\protect\citeauthoryear{{Ball}, {Brunner}, {Myers} \& {Tcheng}}{{Ball}
  et~al.}{2006}]{ball2006robust}
{Ball} N.~M.,  {Brunner} R.~J.,  {Myers} A.~D.,    {Tcheng} D.,  2006, \apj,
  650, 497

\bibitem[\protect\citeauthoryear{{Ben{\'{\i}}tez}}{{Ben{\'{\i}}tez}}{2000}]{benitez2000bayesian}
{Ben{\'{\i}}tez} N.,  2000, \apj, 536, 571

\bibitem[\protect\citeauthoryear{Bertin \& Arnouts}{Bertin \&
  Arnouts}{1996}]{bertin1996sextractor}
Bertin E.,  Arnouts S.,  1996, \aaps, 117, 393

\bibitem[\protect\citeauthoryear{{Bohlin}, {Colina} \& {Finley}}{{Bohlin}
  et~al.}{1995}]{bohlin1995white}
{Bohlin} R.~C.,  {Colina} L.,    {Finley} D.~S.,  1995, \aj, 110, 1316

\bibitem[\protect\citeauthoryear{Breiman}{Breiman}{1996}]{breiman1996stacked}
Breiman L.,  1996, Machine learning, 24, 49

\bibitem[\protect\citeauthoryear{Breiman}{Breiman}{2001}]{breiman2001random}
Breiman L.,  2001, Machine learning, 45, 5

\bibitem[\protect\citeauthoryear{Breiman, Friedman, Stone \& Olshen}{Breiman
  et~al.}{1984}]{breiman1984classification}
Breiman L.,  Friedman J.,  Stone C.~J.,    Olshen R.~A.,  1984, Classification
  and regression trees.
CRC press

\bibitem[\protect\citeauthoryear{Brier}{Brier}{1950}]{brier1950verification}
Brier G.~W.,  1950, Monthly weather review, 78, 1

\bibitem[\protect\citeauthoryear{{Carrasco Kind} \& {Brunner}}{{Carrasco Kind}
  \& {Brunner}}{2013}]{carrascokind2013tpz}
{Carrasco Kind} M.,  {Brunner} R.~J.,  2013, \mnras, 432, 1483

\bibitem[\protect\citeauthoryear{{Carrasco Kind} \& {Brunner}}{{Carrasco Kind}
  \& {Brunner}}{2014a}]{carrascokind2014exhausting}
{Carrasco Kind} M.,  {Brunner} R.~J.,  2014a, \mnras, 442, 3380

\bibitem[\protect\citeauthoryear{{Carrasco Kind} \& {Brunner}}{{Carrasco Kind}
  \& {Brunner}}{2014b}]{carrascokind2014somz}
{Carrasco Kind} M.,  {Brunner} R.~J.,  2014b, \mnras, 438, 3409

\bibitem[\protect\citeauthoryear{{Chabrier}, {Baraffe}, {Allard} \&
  {Hauschildt}}{{Chabrier} et~al.}{2000}]{chabrier2000evolutionary}
{Chabrier} G.,  {Baraffe} I.,  {Allard} F.,    {Hauschildt} P.,  2000, \apj,
  542, 464

\bibitem[\protect\citeauthoryear{{Coleman}, {Wu} \& {Weedman}}{{Coleman}
  et~al.}{1980}]{coleman1980colors}
{Coleman} G.~D.,  {Wu} C.-C.,    {Weedman} D.~W.,  1980, \apjs, 43, 393

\bibitem[\protect\citeauthoryear{Davis et~al.,}{Davis
  et~al.}{2003}]{davis2003science}
Davis M.,  et~al., 2003, Astronomical Telescopes and Instrumentation, pp
  161--172

\bibitem[\protect\citeauthoryear{Erben et~al.,}{Erben
  et~al.}{2013}]{erben2013cfhtlens}
Erben T.,  et~al., 2013, \mnras, p. stt928

\bibitem[\protect\citeauthoryear{Fadely, Hogg \& Willman}{Fadely
  et~al.}{2012}]{Fadely2012}
Fadely R.,  Hogg D.~W.,    Willman B.,  2012, \apj, 760, 15

\bibitem[\protect\citeauthoryear{{Foreman-Mackey}, {Hogg}, {Lang} \&
  {Goodman}}{{Foreman-Mackey} et~al.}{2013}]{Foreman-Mackey2013}
{Foreman-Mackey} D.,  {Hogg} D.~W.,  {Lang} D.,    {Goodman} J.,  2013, \pasp,
  125, 306

\bibitem[\protect\citeauthoryear{Garilli et~al.,}{Garilli
  et~al.}{2008}]{garilli2008vimos}
Garilli B.,  et~al., 2008, \aap, 486, 683

\bibitem[\protect\citeauthoryear{Garilli et~al.,}{Garilli
  et~al.}{2014}]{garilli2014vimos}
Garilli B.,  et~al., 2014, \aap, 562, A23

\bibitem[\protect\citeauthoryear{{G{\'o}rski}, {Hivon}, {Banday}, {Wandelt},
  {Hansen}, {Reinecke} \& {Bartelmann}}{{G{\'o}rski}
  et~al.}{2005}]{gorski2005healpix}
{G{\'o}rski} K.~M.,  {Hivon} E.,  {Banday} A.~J.,  {Wandelt} B.~D.,  {Hansen}
  F.~K.,  {Reinecke} M.,    {Bartelmann} M.,  2005, \apj, 622, 759

\bibitem[\protect\citeauthoryear{Gwyn}{Gwyn}{2012}]{gwyn2012canada}
Gwyn S.~D.,  2012, \aj, 143, 38

\bibitem[\protect\citeauthoryear{{Henrion}, {Mortlock}, {Hand} \&
  {Gandy}}{{Henrion} et~al.}{2011}]{henrion2011bayesian}
{Henrion} M.,  {Mortlock} D.~J.,  {Hand} D.~J.,    {Gandy} A.,  2011, \mnras,
  412, 2286

\bibitem[\protect\citeauthoryear{Heymans et~al.,}{Heymans
  et~al.}{2012}]{heymans2012cfhtlens}
Heymans C.,  et~al., 2012, \mnras, 427, 146

\bibitem[\protect\citeauthoryear{Hildebrandt et~al.,}{Hildebrandt
  et~al.}{2012}]{hildebrandt2012cfhtlens}
Hildebrandt H.,  et~al., 2012, \mnras, 421, 2355

\bibitem[\protect\citeauthoryear{{Kaiser}, {Squires} \& {Broadhurst}}{{Kaiser}
  et~al.}{1995}]{Kaiser1995}
{Kaiser} N.,  {Squires} G.,    {Broadhurst} T.,  1995, \apj, 449, 460

\bibitem[\protect\citeauthoryear{{Kinney}, {Calzetti}, {Bohlin}, {McQuade},
  {Storchi-Bergmann} \& {Schmitt}}{{Kinney} et~al.}{1996}]{kinney1996template}
{Kinney} A.~L.,  {Calzetti} D.,  {Bohlin} R.~C.,  {McQuade} K.,
  {Storchi-Bergmann} T.,    {Schmitt} H.~R.,  1996, \apj, 467, 38

\bibitem[\protect\citeauthoryear{Kohonen}{Kohonen}{1990}]{kohonen1990self}
Kohonen T.,  1990, Proceedings of the IEEE, 78, 1464

\bibitem[\protect\citeauthoryear{Kohonen}{Kohonen}{2001}]{kohonen2001self}
Kohonen T.,  2001, Self-organizing maps.
Vol.~30 of Springer, Springer

\bibitem[\protect\citeauthoryear{{Kron}}{{Kron}}{1980}]{kron1980photometry}
{Kron} R.~G.,  1980, \apjs, 43, 305

\bibitem[\protect\citeauthoryear{Le~F{\`e}vre et~al.,}{Le~F{\`e}vre
  et~al.}{2005}]{le2005vimos}
Le~F{\`e}vre O.,  et~al., 2005, \aap, 439, 845

\bibitem[\protect\citeauthoryear{{Messier}}{{Messier}}{1781}]{messier1781catalogue}
{Messier} C.,  1781, Connoissance des Temps for 1784, pp 227--267

\bibitem[\protect\citeauthoryear{Monteith, Carroll, Seppi \& Martinez}{Monteith
  et~al.}{2011}]{Monteith2011}
Monteith K.,  Carroll J.~L.,  Seppi K.,    Martinez T.,  2011, in Neural
  Networks (IJCNN), The 2011 International Joint Conference on Turning bayesian
  model averaging into bayesian model combination.
pp 2657--2663

\bibitem[\protect\citeauthoryear{Newman et~al.,}{Newman
  et~al.}{2013}]{newman2013deep2}
Newman J.~A.,  et~al., 2013, \ApJSupp, 208, 5

\bibitem[\protect\citeauthoryear{{Odewahn}, {Stockwell}, {Pennington},
  {Humphreys} \& {Zumach}}{{Odewahn} et~al.}{1992}]{odewahn1992automated}
{Odewahn} S.~C.,  {Stockwell} E.~B.,  {Pennington} R.~L.,  {Humphreys} R.~M.,
   {Zumach} W.~A.,  1992, \aj, 103, 318

\bibitem[\protect\citeauthoryear{Paterno}{Paterno}{2003}]{paterno2004calculating}
Paterno M.,  2003, Calculating Efficiencies and Their Uncertainties,
  \url{http://home.fnal.gov/~paterno/images/effic.pdf}

\bibitem[\protect\citeauthoryear{{Pickles}}{{Pickles}}{1998}]{pickles1998stellar}
{Pickles} A.~J.,  1998, \pasp, 110, 863

\bibitem[\protect\citeauthoryear{{Robin} et~al.,}{{Robin}
  et~al.}{2007}]{robin2007stellar}
{Robin} A.~C.,  et~al., 2007, \apjs, 172, 545

\bibitem[\protect\citeauthoryear{Rokach}{Rokach}{2010}]{rokach2010ensemble}
Rokach L.,  2010, Artificial Intelligence Review, 33, 1

\bibitem[\protect\citeauthoryear{Ross et~al.,}{Ross
  et~al.}{2011}]{ross2011ameliorating}
Ross A.~J.,  et~al., 2011, \mnras, 417, 1350

\bibitem[\protect\citeauthoryear{{Sebok}}{{Sebok}}{1979}]{sebok1979optimal}
{Sebok} W.~L.,  1979, \aj, 84, 1526

\bibitem[\protect\citeauthoryear{Sevilla-Noarbe \& Etayo-Sotos}{Sevilla-Noarbe
  \& Etayo-Sotos}{2015}]{sevilla2015effect}
Sevilla-Noarbe I.,  Etayo-Sotos P.,  2015, Astronomy and Computing, in press
  (arXiv:1504.06776)

\bibitem[\protect\citeauthoryear{Silverman}{Silverman}{1986}]{silverman1986density}
Silverman B.~W.,  1986, CRC press, 26

\bibitem[\protect\citeauthoryear{Soumagnac et~al.,}{Soumagnac
  et~al.}{2015}]{soumagnac2013star}
Soumagnac M.~T.,  et~al., 2015, \mnras, 450, 666

\bibitem[\protect\citeauthoryear{{Suchkov}, {Hanisch} \& {Margon}}{{Suchkov}
  et~al.}{2005}]{suchkov2005census}
{Suchkov} A.~A.,  {Hanisch} R.~J.,    {Margon} B.,  2005, \aj, 130, 2439

\bibitem[\protect\citeauthoryear{Swets, Dawes \& Monahan}{Swets
  et~al.}{2000}]{swets2000better}
Swets J.~A.,  Dawes R.~M.,    Monahan J.,  2000, Scientific American, p.~83

\bibitem[\protect\citeauthoryear{Ting \& Witten}{Ting \&
  Witten}{1999}]{ting1999issues}
Ting K.~M.,  Witten I.~H.,  1999, J. Artif. Intell. Res.(JAIR), 10, 271

\bibitem[\protect\citeauthoryear{{Valdes}}{{Valdes}}{1982}]{valdes1982resolution}
{Valdes} F.,  1982, in Instrumentation in Astronomy IV Vol.~331 of Society of
  Photo-Optical Instrumentation Engineers (SPIE) Conference Series, {Resolution
  classifier}.
pp 465--472

\bibitem[\protect\citeauthoryear{{Vasconcellos}, {de Carvalho}, {Gal},
  {LaBarbera}, {Capelato}, {Frago Campos Velho}, {Trevisan} \&
  {Ruiz}}{{Vasconcellos} et~al.}{2011}]{vasconcellos2011decision}
{Vasconcellos} E.~C.,  {de Carvalho} R.~R.,  {Gal} R.~R.,  {LaBarbera} F.~L.,
  {Capelato} H.~V.,  {Frago Campos Velho} H.,  {Trevisan} M.,    {Ruiz}
  R.~S.~R.,  2011, \aj, 141, 189

\bibitem[\protect\citeauthoryear{{Weir}, {Fayyad} \& {Djorgovski}}{{Weir}
  et~al.}{1995}]{weir1995automated}
{Weir} N.,  {Fayyad} U.~M.,    {Djorgovski} S.,  1995, \aj, 109, 2401

\bibitem[\protect\citeauthoryear{Wolpert}{Wolpert}{1992}]{wolpert1992stacked}
Wolpert D.~H.,  1992, Neural networks, 5, 241

\bibitem[\protect\citeauthoryear{{Yee}}{{Yee}}{1991}]{yee1991faint}
{Yee} H.~K.~C.,  1991, \pasp, 103, 396

\bibitem[\protect\citeauthoryear{Yin}{Yin}{2008}]{yin2008self}
Yin H.,  2008, Computational intelligence: a compendium, pp 715--762

\end{thebibliography}
}

\bsp

\label{lastpage}

\end{document}